\documentclass[aps,prb,reprint]{revtex4-1}

\usepackage[a4paper]{geometry} 
\usepackage[utf8]{inputenc} 
\usepackage[T1]{fontenc}  
\usepackage[english]{babel}

\usepackage{tikz}
\usepackage{enumerate}
\usepackage{stmaryrd}
\usepackage{url} 
\usepackage{braket}
\usepackage{tikzorbital}
\usepackage{hyperref}
\usepackage[figure]{hypcap}
\usetikzlibrary{arrows}
\usepackage{amsmath,amsthm,amssymb,amsfonts,extarrows} 
\usepackage{bbm} 
\usepackage{graphicx}
\usepackage{epstopdf}
\usepackage{subfigure}
\usepackage{setspace}
\usepackage{threeparttable}
\usepackage{enumitem}
\usepackage{booktabs}
\usepackage{float}
\usepackage{siunitx}
\usepackage{framed}
\usepackage{natbib} 
\usepackage{dashrule}

\theoremstyle{plain} 
\newtheorem{theorem}{Theorem}[section]

\newtheorem{lemma}[theorem]{Lemma}
\newtheorem*{theoremWithoutNumber}{Theorem}

\theoremstyle{definition} 
\newtheorem{definition}[theorem]{Definition}

\usepackage{mathtools}
\usepackage{letltxmacro}

\LetLtxMacro\orgvdots\vdots
\LetLtxMacro\orgddots\ddots

\makeatletter
\DeclareRobustCommand\vdots{%
	\mathpalette\@vdots{}%
}

\DeclareRobustCommand\vdots{%
	\mathpalette\@vdots{}%
}
\newcommand*{\@vdots}[2]{%
	\sbox0{$#1\cdotp\cdotp\cdotp\m@th$}%
	\sbox2{$#1.\m@th$}%
	\vbox{%
		\dimen@=\wd0 %
		\advance\dimen@ -3\ht2 %
		\kern.5\dimen@
		\dimen@=\wd2 %
		\advance\dimen@ -\ht2 %
		\dimen2=\wd0 %
		\advance\dimen2 -\dimen@
		\vbox to \dimen2{%
			\offinterlineskip
			\copy2 \vfill\copy2 \vfill\copy2 %
		}%
	}%
}
\DeclareRobustCommand\ddots{%
	\mathinner{%
		\mathpalette\@ddots{}%
		\mkern\thinmuskip
	}%
}
\newcommand*{\@ddots}[2]{%
	\sbox0{$#1\cdotp\cdotp\cdotp\m@th$}%
	\sbox2{$#1.\m@th$}%
	\vbox{%
		\dimen@=\wd0 %
		\advance\dimen@ -3\ht2 %
		\kern.5\dimen@
		\dimen@=\wd2 %
		\advance\dimen@ -\ht2 %
		\dimen2=\wd0 %
		\advance\dimen2 -\dimen@
		\vbox to \dimen2{%
			\offinterlineskip
			\hbox{$#1\mathpunct{.}\m@th$}%
			\vfill
			\hbox{$#1\mathpunct{\kern\wd2}\mathpunct{.}\m@th$}%
			\vfill
			\hbox{$#1\mathpunct{\kern\wd2}\mathpunct{\kern\wd2}\mathpunct{.}\m@th$}%
		}%
	}%
}

\begin{document}
	\parindent0pt 
	\title{Complex Ground-State and Excitation Energies in Coupled-Cluster Theory}
	\date{\today}
	\author{Simon Thomas, Florian Hampe, Stella Stopkowicz, and Jürgen Gauss}
	\affiliation{Department Chemie, Johannes Gutenberg-Universität Mainz, Duesbergweg 10-14, 55128 Mainz, Germany}

%	Die Coupled-Cluster Theorie hat sich in den vergangenen Jahren zum meist-verwendeten Verfahren für genaue, quantenchemische Rechnungen entwickelt. Um Anregungsenergien zu berechnen wird meist die darauf aufbauende Equation-of-Motion Coupled-Cluster Theorie genutzt. Diese nicht hermitesche Theorie liefert in der Praxis in vielen Fällen sehr gute Näherungswerte.  Aufgrund der fehlenden Hermitizität kann es jedoch in manchen Fällen zu komplexen Werten kommen, was zunächst einmal dem natürlichen Verständnis der Physik widerspricht. \\
%	In der vorliegenden Arbeit werden anfangs einige mathematischen Ergebnisse aus dem Bereich der Eigenwerttheorie hergeleitet.
%	Anschließend sollen die Grundlagen der Coupled-Cluste Theorie, sowie der EOM-CC Theorie dargestellt werden, sodass die Ergebnisse aus der Eigenwerttheorie in diesen Bereich übertragen werden können. Man stellt fest, dass die Szenarien, die bei EOM-CC Rechnungen auftreten, zum größten Teil im Rahmen allgemeiner Eigenwerttheorie erklärbar sind. Dazu gehören beispielsweise, dass in vielen Fällen die erhaltenen Werte trotz fehlender Hermitizität reell bleiben, andererseits, dass das Auftreten von komplexen Eigenwerten in der Nähe von sogenannten 'konischen Schnitten' kritisch ist. \\
%	Abschließend werden ein paar Ideen für alternativen Methoden vorgestellt, die das Problem der komplexen Eigenwerte teilweise oder ganz umgehen.
%	
\begin{abstract}
	Since in coupled-cluster (CC) theory ground-state and excitation energies are eigenvalues of a non-Hermitian matrix, these energies can in principle take on complex values.
    	In this paper we discuss the appearance of complex energy values in CC calculations from a mathematical perspective. We analyze the behaviour of the eigenvalues of Hermitian matrices that are perturbed (in a non-Hermitian manner) by a real parameter. Based on these results we show that for CC calculations with real-valued Hamiltonian matrices the ground-state energy generally takes a real value. Furthermore, we show that in the case of real-valued Hamiltonian matrices complex excitation energies only occur in the context of conical intersections. In such a case, unphysical consequences are encountered
    	such as a wrong dimension of the intersection seam, large numerical deviations from full configuration-interaction (FCI) results, and the square-root-like behaviour of the potential surfaces near the conical intersection. 
    	In the case of CC calculations with complex-valued Hamiltonian matrix elements, it turns out that complex energy values are to be expected  for ground and excited states when no symmetry is present. We confirm the occurrence of complex energies by sample calculations using a six-state model and by CC calculations for the $\mathrm{H_2O}$ molecule in a strong magnetic field. We furthermore show that symmetry can prevent the occurrence of complex energy values. Lastly, we demonstrate that in most cases the real part of the complex energy values provides a very good approximation to the FCI energy.
\end{abstract}
	\maketitle
	%\tableofcontents 

		\section{Introduction}
		\label{sec: Introduction}
		%1.Teil: Ziel: Autor von Wichtigkeit des Themas überzeugen:
		%Inhalt: CC- ist toll, denn Größenkonsistent, etabliert und mit EOM-CC gibts sogar Anregungsenergien
		%Komplexe Werte sind nicht toll, da sie zunächst einmal der Physik widersprechen
		%Komplexe Werte müssen richtig eingeordnet werden. 
		%\note{This paragraph is to briefly explain CC and to explain the problem}\\
		Coupled-cluster (CC) theory\cite{SB09} is one of the most widely used quantum-chemical methods for high-accuracy computations of energies and properties. As a post-Hartree-Fock method, CC theory focuses on an adequate, i.e., size-extensive, treatment of electron correlation and ensures this by applying the exponential of an excitation operator, i.e., the so-called cluster operator, to a reference determinant, most often 
		%but not necessarily 
		chosen as the Hartree-Fock (HF) wave function. The equation-of-motion CC 
		(EOM-CC) ansatz\cite{Emrich81,Stanton93,Comeau93,Rico93,SB09} extends ground-state CC theory to excited states. % that excitation energies can be obtained as well. 
		The key step lies in the similarity transformation of the electronic Hamiltonian with the exponential of the cluster operator followed by a diagonalization of the resulting effective Hamiltonian. However, as this transformation is not unitary, Hermiticity is lost and as a consequence complex excitation energies can in principle be obtained in an EOM-CC calculation.
		
		Hättig\cite{Haettig2005} was the first to note that the lack of Hermiticity can lead in EOM-CC calculations to a qualitatively wrong description of potential energy surfaces in the vicinity of conical intersections.  Using a two-state model, H{\"a}ttig predicted that the energies of the two involved states pass through a point of degeneracy and then enter an area where their values are complex.		
		 Köhn and Tajti\cite{Koehn2007} confirmed this scenario based on EOM-CC calculations for two excited states of formaldehyde ($\mathrm{CH_2O}$). In addition, they observed a square-root like behaviour of the EOM-CC energies of the two states near the intersection and showed that the eigenvectors associated with the two degenerate states become linearly dependent. Kjønstadt {\it et al.}\cite{KMMK2017} later demonstrated that the EOM-CC description of a conical intersection is not necessarily always flawed but depends on whether the similarity-transformed Hamiltonian matrix is defective or not at the point of degeneracy. A qualitatively  correct description is only observed in the case of a non-defective matrix.
		We also note that complex energies have so far not been observed in ground-state CC calculations.

		In this paper we explain why complex energies have not been observed in CC calculations except close to conical intersections
		and discuss in which cases they can be expected.
		%references for previous studies
		%Unlike previous studies,\cite{Haettig2005,Koehn2007,KMMK2017} 
		We %derive our findings by analyzing 
		analyze the behavior of the eigenvalues of general real and complex matrices and apply the corresponding mathematical tools to CC theory. 
		Apart from results that are already discussed in the literature, this approach also leads to 
		%On the one hand we obtain results that have already been presented in the literature. On the other hand, we gain 
		additional knowledge about the shape of the potential surfaces near conical intersections and about the occurrence of complex eigenvalues in the case of Hamiltonian matrices with complex-valued entries. The latter allows us to draw conclusions about the occurrence of complex energy values in the case of CC calculations for systems in a finite magnetic field\cite{S15,HS17,HGS20} and for relativistic CC calculations that include spin-orbit coupling.\cite{Visscher96,WGW08,Shee18,Liu18} \\
		%4. Teil Ziel Überblick über die Kapitel geben
		% 1. theoretische Grundlagen in chapter 1. Darin wie sich die Eigenwerte einer Matrix unter Störung der Matrixeinträge verhalten. Und kurze Skizze der coupled-cluster theory.
		% 2. Analyse des Auftretens von komplexen Eigenwerten im Falle einer reellen Matrix. Es wird deutlich, dass das nur im Zusammenhang von konischen Überschneidungen passiert. Konsequenzen des Auftretens der komplexen Werte wie falsche Dimension des intersections seams und des shapes der konischen Überschneidung werden mit den vorherigen mathematischen Ergebnissen erklärt.
		% 3. Analyse des Auftretens von komplexen Energiewerten im Falle einer komplexen Matrix. Es wird klar, dass komplexe Energiewerte zu erwarten sind.Kurzer Abriss wo dies in der Praxis auftaucht und  Illustrierendes Beispiel: Wasser-Molekül im Magnetfeld
		The present work begins with a discussion of a several mathematical definitions and theorems needed for our investigation, like  %some basic mathematical facts, in particular concerning 
		the perturbative analysis of the eigenvalues of a matrix. %Afterwards,
		In section~\ref{sec: Theory}, the basics of CC theory and EOM-CC theory are briefly reviewed. Section~\ref{sec: Complex energy} analyzes the occurrence of eigenvalues in the case of a real-valued Hamiltonian matrix. It is shown that complex energy values can only occur in the context of conical intersections. 
		Using the mathematical tools presented in section~\ref{sec: Theory}, consequences of the occurrence of complex values such as a wrong dimension of the intersection seam and a wrong shape of the potential energy surfaces around the conical intersection are derived and analyzed.
		%Consequences of the occurrence of complex values such as a wrong dimension of the intersection seam and a wrong shape of the conical intersection are obtained with the mathematical tools presented in section~\ref{sec: Theory}.
		Section~\ref{sec::complexValuesComplHam} examines the occurrence of complex energy values in the case that the Hamiltonian matrix has complex-valued entries, as it happens in the case of finite magnetic-field and relativistic CC calculations. Here, we perform and discuss example calculations that show that the appearance of complex energy values is common.%the normal case. 
		We also show that the real part of a complex energy value nevertheless provides a useful approximation to the actual energy. Finally, we demonstrate that symmetry typically ensures %can in some cases ensure 
		that the resulting CC energy values are real.%, which is the main reason why complex energies in the context of finite-field and relativistic CC calculations have not been noticed before.  
		%%%%%%%%%%%%%%%%
		%ANMERKUNG Stella: Ich hab den letzten Satz rausgenommen, denn ich bin mir nicht so sicher, dass das für relativistische Rechnungen stimmt, ich denke eher hier liegt es daran, dass einfach per-default garnicht der Imaginärteil berechnet/gedruckt wird.
		%%%%%%%%%%%%%%%%

	\section{Theory}
Subsection~\ref{subsec: Mathematical Background} reviews the required mathematical background of eigenvalue theory and subsection~\ref{subsec: Coupled cluster} outlines CC theory. 
\label{sec: Theory}
\subsection{Mathematical background}
			\label{subsec: Mathematical Background}
			In the following part two basic series expansions, the well-known Taylor series and the lesser known Puiseux series, are %of relevance.
			discussed.\cite{Wilkinson1965, Kato, Wall2004, Thomas18}
			\begin{definition}
				A formal series expansion 
				\begin{align*}
				y(x)=y_0+a_1x+a_2x^2+a_3x^3+\cdots,
				\end{align*} 
				where $a_i\in \mathbb{C}, \hspace*{0.2cm}  x\in \mathbb{R},$
				is denoted as a \emph{Taylor series} at point $x=0$.\\
				A formal series expansion of the form
				\begin{align*}
					y(x)=y_0+a_1(x^\frac{1}{m})_k+a_2(x^\frac{1}{m})_k^2+\cdots,
				\end{align*}
				where $m\in \mathbb{N}$, $m>1$, $a_i\in \mathbb{C}$ and $x\in \mathbb{R}$,
				is called a \emph{Puiseux series} at point $x=0$.\\
				Here $(x^\frac{1}{m})_k:=|x|^\frac{1}{m} e^{i\frac{2 \pi k+\pi}{m}}$ is an $m$-th root of $x$.  
			\end{definition}
		It is obvious that different choices for the root $(x^\frac{1}{m})_k$ lead to different branches of the series. All branches are continuous and analytic in all points except for $x=0$.\\
		We call a matrix analytically dependent on the parameter $\epsilon$ if all matrix entries can be described by a Taylor series depending on $\epsilon$. A change of the matrix entries due to a variation of $\epsilon$ is called an analytic matrix perturbation. The following theorem\cite{Kato, Thomas18} states that the change of the eigenvalues caused by an analytic matrix perturbation can be described either by a Taylor series or by a Puiseux series.

\begin{figure}
	 			\label{fig: Real and}
	 			\centering
	 			\includegraphics[trim = 20mm 75mm 20mm 60mm, angle=0, clip, width=0.45\textwidth]{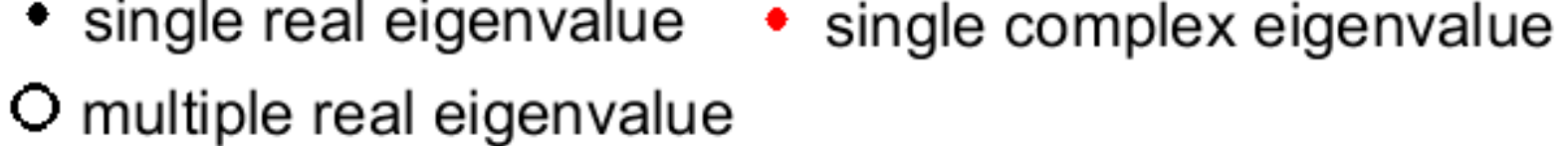}\\
				\includegraphics[width=0.44\textwidth]{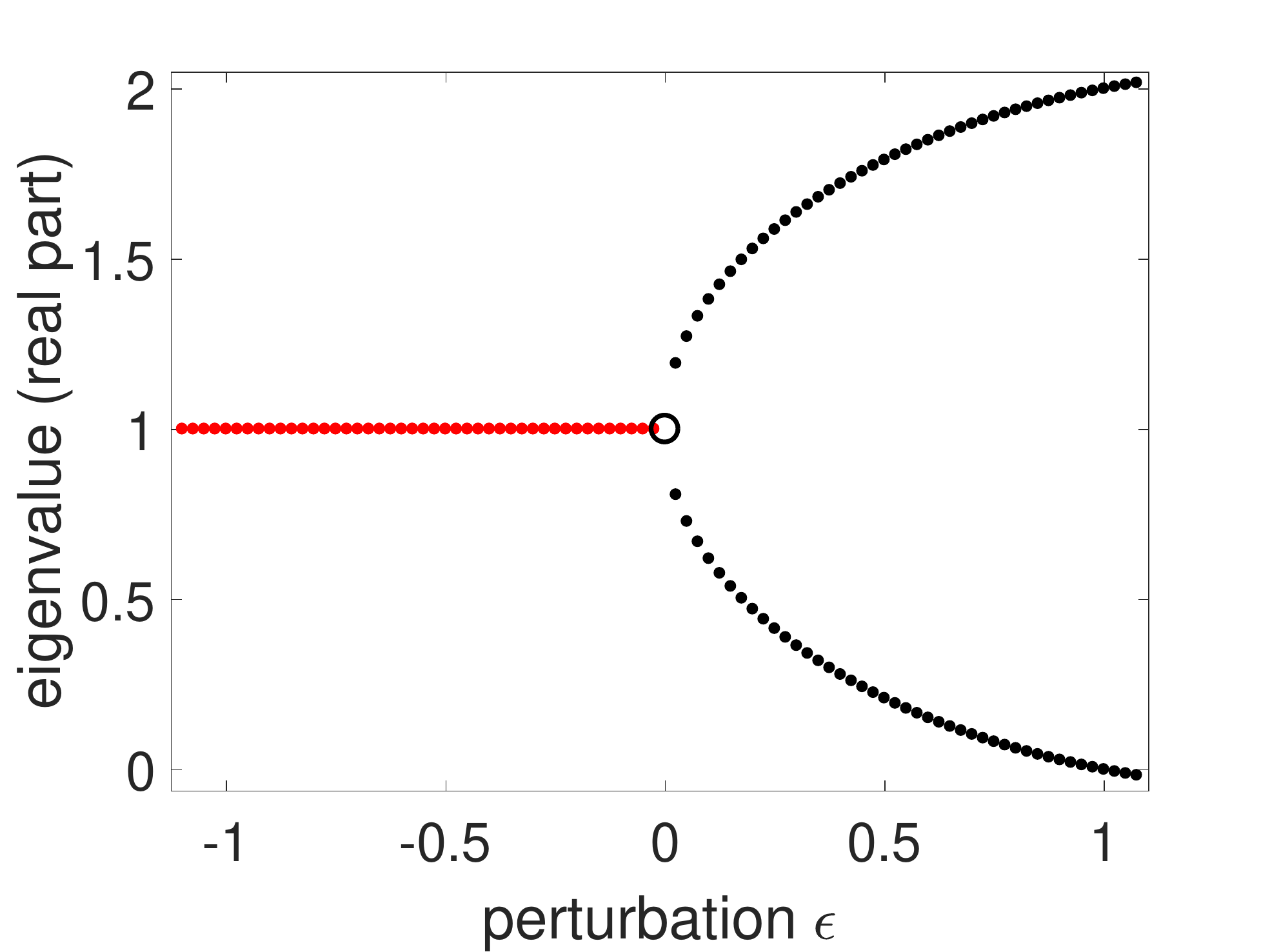}
				 \includegraphics[width=0.45\textwidth]{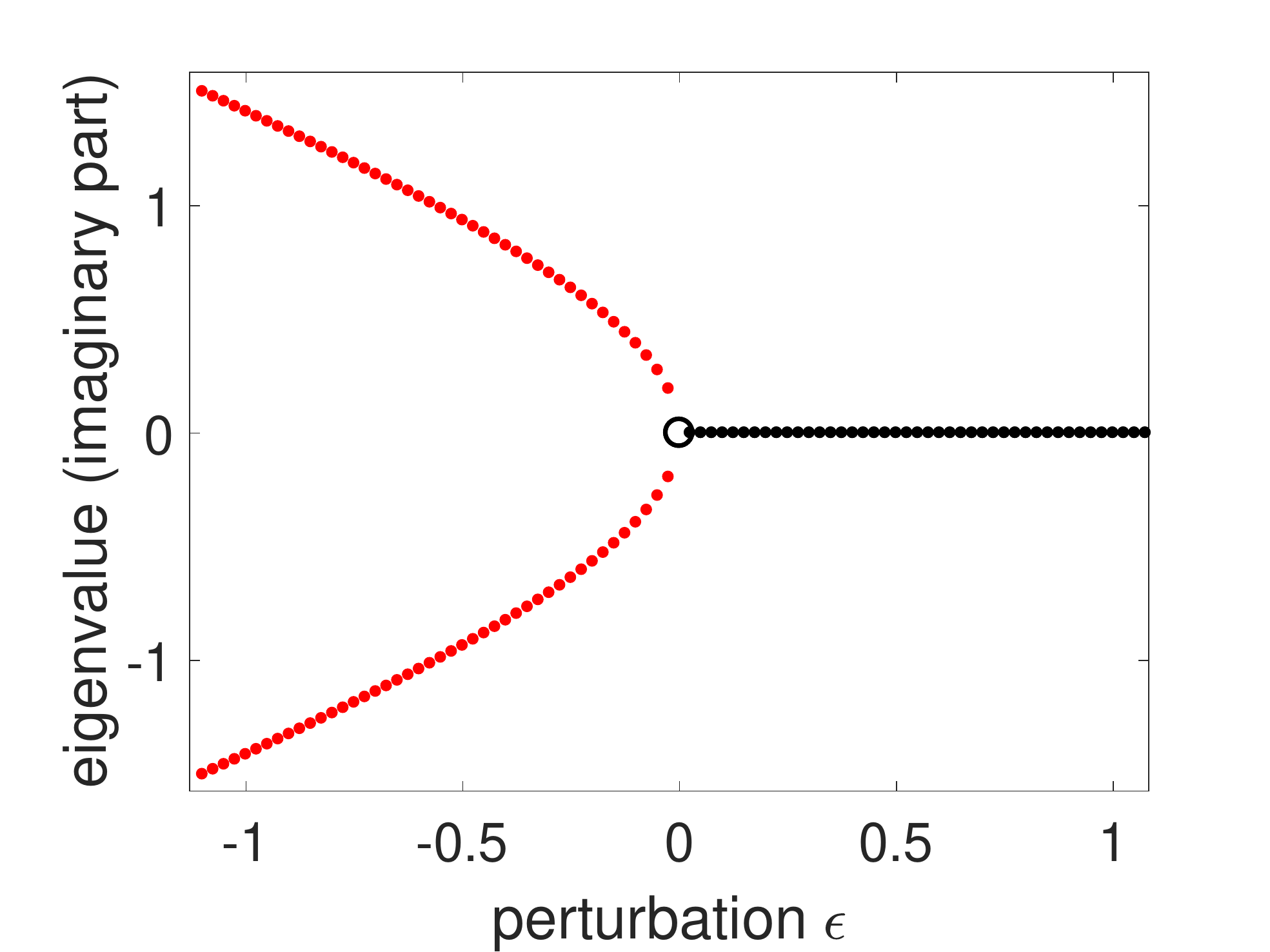}\\
	 			\caption{Real and imaginary part of the eigenvalues of $A_1(\epsilon)$.}
	 		\end{figure}

		\begin{theorem}
			\label{theorem: locale series}
			Let $A(\epsilon) \in \mathbb{C}(n \times n)$ be a matrix whose entries depend analytically on one real parameter $\epsilon$. Furthermore, let $\lambda_i(\epsilon)$ be an eigenvalue of $A(\epsilon)$. Then the following holds:
			\begin{itemize}
				\item Let $\lambda_i(0)$ be a single eigenvalue. Then there exists a neighbourhood $U$ of $(0,\lambda_i(0))$, where exactly one single eigenvalue exists for every $\epsilon \in U$.  The dependence of the eigenvalue on $\epsilon$ can be expressed by means of a Taylor series as
				\begin{align}
				\label{equation: potential series}
				\lambda_i(\epsilon)=\lambda(0)+p_{i1}\epsilon+p_{i2}\epsilon^2+p_{i3}\epsilon^3+\cdots.
				\end{align}
				\item  Let $\lambda(0)=\lambda_1(0)=\lambda_2(0)=...=\lambda_m(0)$ be a multiple eigenvalue. Then there exists a neighbourhood $U$ of $(0,\lambda(0))$, such that for each fixed $\epsilon$ there exist exactly $m$ eigenvalues of $A(\epsilon)$ in $U$. The dependence of each eigenvalue on $\epsilon$ can be described either by a Taylor series (as in Eq.~(\ref{equation: potential series})) or by one of the branches, $\lambda_i(\epsilon)$, of a Puiseux series $\lambda(\epsilon)$, which has the form 
				\begin{align}
					\label{equation: puiseux series of lambda}
					\lambda(\epsilon)=\lambda(0)+p_{i1}\epsilon^\frac{1}{\widetilde{m}}+p_{i2}\epsilon^\frac{2}{\widetilde{m}}+\cdots
				\end{align}
				with $\widetilde{m}\leq m$.
				In the case that one of the branches of a Puiseux series is a solution of the eigenvalue problem $A(\epsilon)v(\epsilon)=\lambda(\epsilon)v(\epsilon)$, the other possible branches $\lambda_i(\epsilon)$ fulfill the eigenvalue equation, too.
			\end{itemize}
		\end{theorem}
	 	We illustrate the stated theorem by analyzing the behaviour of the eigenvalues of the following $2\times 2$ matrix  
	 		\begin{align*}
		 		A_1(\epsilon)=\begin{pmatrix}
		 		1-\epsilon & 1-\epsilon \\
		 		1.5\epsilon & 1+\epsilon
		 		\end{pmatrix}
	 		\end{align*}
	 		which analytically depends on a real parameter $\epsilon$.
	 		For $\epsilon \neq 0$ the matrix $A(\epsilon)$ has single eigenvalues. They develop in analytic manner as a function of the perturbation parameter (see theorem \ref{theorem: locale series} and Figure \ref{fig: Real and}).
	 		For $\epsilon=0$ a multiple eigenvalue occurs.
	 		The series expansion at this point is
	 		\begin{align}
	 			\label{formula: puiseux series}
		 		\lambda(\epsilon)\approx 1 + 1.225\epsilon^{\frac{1}{2}}-0.204{\epsilon}^\frac{3}{2}-0.017{\epsilon}^\frac{5}{2}-\cdots .
	 		\end{align}

 		According to this series expansion the eigenvalues develop in a continuous manner. 
 		From the same series expansion as well as from Figure \ref{fig: Real and} we also see that the square-root term dominates the behaviour of the eigenvalues near the branching point $\epsilon=0$. In case of such a shape we speak of \emph{square-root like behaviour} in the broader context. Consequences of the square-root like behaviour are that the function $\lambda(\epsilon)$ is not differentiable at the point $\epsilon=0$ and that a small change in $\epsilon$ leads to a large change in the eigenvalues. \\
 		In the given example, the eigenvectors of the matrix $A_1(0)$ are linearly dependent. Such a matrix is called defective, whereas a matrix is called non-defective if its eigenvectors span a complete base of the vector space (i.e., the matrix is diagonalizable).\cite{Golub2013}
		In case that the matrix is non-defective at the point where multiple eigenvalues occur a square-root like behaviour of the eigenvalues cannot appear. This is ensured by the following theorem by Kato:\cite{Kato}	
 	\begin{theorem}[Kato]
 		\label{theorem: smooth path}
 		Let $A(\epsilon) \in \mathbb{C}(n \times n)$ be a non-defective matrix depending analytically on $\epsilon$. Let $\lambda(0)=\lambda_1(0)=\lambda_2(0)=...=\lambda_m(0)$ be a multiple eigenvalue.  Then, each eigenvalue $\lambda_i(\epsilon)$ in the neighbourhood of $(0,\lambda(0))$ can be represented in one of the following two ways:
 		\begin{itemize}
 			\item
 			by a Taylor series 
 			\begin{align*}
 			\lambda_i(\epsilon)=\lambda(0)+p_{i1}\epsilon+p_{i2}\epsilon^2+p_{i3}\epsilon^3+\cdots .
 			\end{align*}
 			\item
 			or by a branch of a Puiseux series, where the linear term dominates the series expansion:				
 			\begin{align*}
 			\lambda(\epsilon)=\lambda(0)+p_{i1}\epsilon + p_{i2}\epsilon^{1+\frac{1}{\widetilde{m}}}+p_{i3}\epsilon^{1+\frac{2}{\widetilde{m}}}+\cdots
 			\end{align*}
  		\end{itemize}
  				with $\widetilde{m}\leq m$. In both cases $\lambda_i(\epsilon)$ is differentiable at the point $\epsilon=0$. Note that the second case equals the first case if $\widetilde{m}=1$.
 	\end{theorem}
Thus, the question whether or not a matrix is defective plays a decisive role in the appearance of the function $\lambda(\epsilon)$.\\
 
 In the previous example, the transition from complex to real eigenvalues, caused by the variation of $\epsilon$, proceeds via a multiple eigenvalue.
 This statement is generally valid and is explained in detail by means of the following theorem. 
	\begin{theorem}
		\label{theorem: multiple eigenvalues}
		Let $\lambda_i(\epsilon)$ be an eigenvalue of $A(\epsilon) \in \mathbb{C}(n \times n)$. Furthermore, there exist $a,b>0$ such that $\lambda_i(\epsilon)$ takes a real value for all $\epsilon \in [-a,0]$  and $\lambda_i(\epsilon)$ takes a complex value for all $\epsilon \in (0,b]$. Then:
		\begin{enumerate}[label=(\alph*)]
			\item In a neighbourhood of $\epsilon=0$ the eigenvalue $\lambda_i(\epsilon)$ can be represented by a branch of a Puiseux series.
			\item For $\epsilon=0$ a multiple eigenvalue occurs.
		\end{enumerate} 
	\end{theorem}
	A proof can be found in the Appendix \ref{Mathematical additions}.

	In the case that the matrix $ A(\epsilon) $ has only real entries, further properties can be specified for the behaviour of an eigenvalue $\lambda(\epsilon)$:

	\begin{theorem}
		\label{theorem: real neigborhood}
			Let $A(\epsilon)$ be a matrix with only real entries for all $\epsilon\in (-r,r)$, then:
			\begin{itemize}
				\item Let $\lambda(\epsilon)$ be a complex eigenvalue of $A(\epsilon)$, then the complex-conjugated eigenvalue ${\lambda^*(\epsilon)}$ is also an eigenvalue of $A(\epsilon)$.
				\item Let $\lambda_i(0)$ be a single real eigenvalue of $A(0)$, then a neighbourhood $U$ of $(0,\lambda(0))$ exists, such that $\lambda_i(\epsilon)$ takes only real values in $U$.
			\end{itemize}
	\end{theorem}
	The proof can also be found in the Appendix \ref{Mathematical additions}.

	\subsection{Coupled-cluster theory}
	\label{subsec: Coupled cluster}
	%\note{Schrödinger-equation}\\
	The electronic states $\ket{\Psi_k}$ of a molecule together with their associated energy values $E_k$ are determined by the electronic Schrödinger equation:
	\begin{equation*}
	\label{equation: Schroedinger equation}
	\hat{H}\ket{\Psi_k}=E_k\ket{\Psi_k}, \hspace{0.5cm}k\ge0.
	\end{equation*}
	
	In second quantization, the Hamiltonian $\hat{H}$ takes the form\cite{SB09}
	\begin{equation}
	\label{equation: Hamiltonian}
	\hat{H} = \sum_{p,q}h_{pq}\hat{a}_p^\dagger \hat{a}_p + \frac{1}{4}\sum_{p,q,r,s}g_{pqrs}\hat{a}_p^\dagger \hat{a}_q^\dagger \hat{a}_s \hat{a}_r
	\end{equation}
	with $\{p,q, \cdots\}$ representing the index set of the underlying molecular spin orbitals $\{\varphi_p, \varphi_q, ...\}$ and  $\hat{a}^\dagger_p$ and $\hat{a}_p$ as the corresponding %particle
	elementary creation and annihilation operators.	
	In Eq.~(\ref{equation: Hamiltonian}), $h_{pq}$ and $g_{pqrs}$ denote the matrix elements of the one ($\hat{h}$)  and two-electron ($\hat{g}$) operator which constitute the Hamiltonian in first quantization. They are calculated using the underlying set of one-electron functions $\{\varphi_p, \varphi_q, ...\}$ and the operators $\hat{h}$ and $\hat{g}$ via
	\begin{align}
	h_{pq}&=\bra{\varphi_p}\hat{h}\ket{\varphi_q}\\
	g_{pqrs}&=\bra{\varphi_p\varphi_q}\hat{g}\ket{\varphi_r\varphi_s}-\bra{\varphi_p\varphi_q}\hat{g}\ket{\varphi_s\varphi_r}.
	\end{align}
	The exact definitions of the one- and two-electron operators $\hat{h}$ and $\hat{g}$ depend on the context. They are different for traditional nonrelativistic CC calculations and those that incorporate relativistic effects,\cite{Faegri} and for cases in which a finite magnetic field is present.\cite{S15, HS17,HS19,HGS20}	
	For our discussion the general representation given in Eq.~(\ref{equation: Hamiltonian}) is sufficient. However, it must be noted that the choice of the operators $\hat{h}$ and $\hat{g}$ plays a decisive role as to whether the matrix elements $h_{pq}$ and $g_{pqrs}$ are real- or complex-valued.\\
	
%	\note{Ansatz for CC-Function and CC-Equations}\\
	In CC theory,\cite{BM07, SB09, Schneider2009} the ground-state wave function is obtained by applying the exponential of the cluster operator ${\hat{T}}$ to a reference Slater determinant $\psi_0$:
	\begin{equation}
	\label{equation: exponential ansatz}
	\ket{\Psi_{\mathrm{CC}}} = e^{\hat{T}}\ket{\psi_0}.
	\end{equation}
	In second quantization, the cluster operator $\hat{T}$ is given as
	\begin{align}
	\hat{T}&=\hat{T}_1 + \hat{T}_2 +\cdots + \hat{T}_n\\
	&= \sum_{n=1}^{N}\Big(\frac{1}{n!}\Big)^2\sum_{i,j,...}\ \sum_{a,b,...}t_{ij...}^{ab...}\hat{a}_a^\dagger \hat{a}_i \hat{a}_b^\dagger \hat{a}_j \cdots.
	\end{align}
    with the number of electrons $N$ and $i,j,k,\dots$ as well as $a,b,c,\dots$ referring to the occupied and virtual space, respectively.  
	The amplitudes $t_{ij...}^{ab...}$ of the cluster operator are obtained by solving the CC equations
	\begin{equation}
	\label{equation: CC-equations}
	\bra{\psi_I}e^{\hat{T}}\hat{H}e^{-\hat{T}}\ket{\psi_0}=0
	\end{equation}
	for all Slater determinants $\psi_I$ of the FCI space. 
Under certain conditions, which are in particular that the Slater determinants $\psi_I$ as well as the matrix elements $h_{pq}$ and $g_{pqrs}$ are real-valued, this non-linear system of equations has a locally unique real solution for the amplitudes $t_{ij...}^{ab...}$. \cite{Schneider2009} In this case the CC ground-state energy $E_{\mathrm{CC}}=\bra{\psi_0}e^{-\hat{T}}\hat{H}e^{\hat{T}}\ket{\psi_0}$ is also real. 

For computational reasons the cluster operator is usually truncated after a few terms and the CC equations (Eq.~(\ref{equation: CC-equations})) are solved only for the excitations included in $\hat{T}$. For example, by choosing $\hat{T}=\hat{T}_1+\hat{T}_2$ and by solving the CC equations for single and double excitations, the well-known CC singles and doubles (CCSD) method\cite{Purvis82} results. The statement that the system of equations (Eq.~(\ref{equation: CC-equations})) has a locally unique real solution, (if the same conditions hold as before) is also correct for a truncated operator $\hat{T}$.

%\note{Ansatz for EOM-CC-Function and Eigenvalue equation}
	Based on CC theory, the EOM-CC approach\cite{ROW68, Emrich81, Stanton1993,Comeau93,Rico93} is a popular choice for the calculation of excitation energies with the following ansatz for the corresponding $k$-th excited-state wave function:
	\begin{equation}
		\ket{\Psi_{\mathrm{exc}}^{(k)}}=\hat{R}^{(k)}\ket{\Psi_{\mathrm{CC}}}=\hat{R}^{(k)}e^{\hat{T}}\ket{\psi_0}.
	\end{equation} The cluster operator $\hat{T}$ is taken from a preceding CC ground-state calculation and $\hat{R}^{(k)}$ is a linear excitation operator that differs from $\hat{T}$ only by the constant contribution $\hat{R}_0^{(k)}$:
	\begin{equation}
	\hat{R}^{(k)}=\hat{R}^{(k)}_0+\hat{R}^{(k)}_1+\hat{R}^{(k)}_2+\cdots.
	\end{equation} 
	Both the amplitudes of the excitation operator $\hat{R}^{(k)}$, as the entries in the eigenvector $\vec{r}^{(k)}$, and the excited-state energies $E_k$ are obtained by solving the eigenvalue problem
	\begin{equation}
		\bar{H}_{\mathrm{FCI}}\vec{r}^{(k)}= E_k\vec{r}^{(k)},
	\end{equation}
where $\bar{H}_{\mathrm{FCI}}$ is the matrix representation of the similarity-transformed Hamilton operator $\hat{\bar{H}}=e^{\hat{T}}\hat{H}e^{-\hat{T}}$ in the FCI space.
	
To render EOM-CC calculations feasible the excitation operator $\hat{R}^{(k)}$ is usually truncated at the same level as $\hat{T}$. The amplitudes of $\hat{R}^{(k)}$ and the energy values are then determined by the eigenvectors and eigenvalues of the truncated matrix
	\begin{equation}
		\bar{H}_\mathrm{P}=P\bar{H}_{\mathrm{FCI}}P^{\dagger}=\begin{pmatrix}
			E_{\mathrm{CC}} & \hspace*{0.3cm}  &* \\
			0 \\
			\vdots &  & ** \\
			0 
		\end{pmatrix},
	\end{equation} 
	where the matrix $P=\left(\begin{smallmatrix}
	1 & \cdots & 0 & \cdots& 0\\
	\vdots & \ddots & \vdots & & \vdots\\
	0 &\cdots & 1 & \cdots & 0 \\
	\end{smallmatrix}\right)$ projects $\bar{H}_{\mathrm{FCI}}$ onto the space of Slater determinants considered by $\hat{R}^{(k)}$.
	
	Since the first column, apart from the first entry, vanishes due to the CC equations, the CC ground-state energy $E_{\mathrm{CC}}$ equals the lowest eigenvalue $E_0$ of the matrix $\bar{H}_\mathrm{P}$. Thus, the analysis of both the CC ground-state energy $E_{\mathrm{CC}}$ as well as the EOM-CC excitation energies $E_k$ can be performed by means of an analysis of the eigenvalues of $\bar{H}_\mathrm{P}$.\\
	A frequently used EOM-CC scheme is to choose $\hat{T}= \hat{T}_1 + \hat{T}_2$ and $\hat{R}= \hat{R}_0 + \hat{R}_1 + \hat{R}_2$ which leads to
the EOM-CCSD model.\cite{Stanton1993,Comeau93,Rico93} In the present context
this model will be used as a representative for all EOM-CC methods.
	In contrast to Hermitian quantum-chemical methods (e.g., FCI or truncated configuration interaction (CI)), it cannot be ensured that the energy values of the EOM-CC method are real-valued,\cite{Haettig2005,Koehn2007} as the matrix $\bar{H}_\mathrm{P}$ is usually not Hermitian. \\

		\section{Complex energies in case of a real-valued $\bar{H}_\mathrm{P}$ matrix}
		\label{sec: Complex energy}
		According to the previous section, it depends on the matrix $\bar{H}_\mathrm{P}$  whether the EOM-CC energy values (including the CC ground-state energy) are real or not. For the sake of simplicity we limit our discussion to %only discuss in the following 
		the EOM-CCSD model. Conceptually, analogous results can be obtained for other EOM-CC methods (e.g., EOM-CCSDT,\cite{eomccsdt1,eomccsdt2,eomccsdt3} EOM-CCSDTQ,\cite{eomccsdtq} etc.) by means of a similar analyses. %analogous results can be derived by means of a similar analysis. 
		In this section it is assumed that the entries of the matrix $\bar{H}_\mathrm{FCI}$ are real. This applies, for example, if the matrix elements $h_{pq}$, $g_{pqrs}$ and the underlying one-electron wave functions $\{\varphi_p, \varphi_q, ...\}$ are real-valued (see section \ref{subsec: Coupled cluster}).
		As a first approach to analyze the appearance of complex energy values, we construct a continuous connection between the FCI and EOM-CCSD energy values.

		\subsection{Connection between FCI and EOM-CCSD energy values}
		\label{subsec: Connection between}
	
		Let $\bar{H}_{\mathrm{FCI}}$ be the matrix representation of the similarity transformed Hamiltonian in the FCI space and $\bar{H}_\mathrm{P}=P\bar{H}_{\mathrm{FCI}}P^\dagger$ the truncated $\bar{H}_{\mathrm{FCI}}$ matrix as described in section \ref{subsec: Coupled cluster}.
		The eigenvalues of $\bar{H}_\mathrm{FCI}$, hereafter also referred to as FCI eigenvalues, are the energy eigenvalues of the FCI method. The eigenvalues of the matrix $\bar{H}_\mathrm{P}$, hereafter referred to as CCSD eigenvalues, are the energy values of the EOM-CCSD method.
		We now establish a continuous connection between the FCI and CCSD eigenvalues by switching on a perturbation $S$ using a real parameter $\varepsilon$.
		Formally, this can be described by
		\begin{align*}
			\bar{H}(\varepsilon):=\bar{H}_\mathrm{FCI}+ \varepsilon
			\underbrace{
			\begin{pmatrix}
				0 & * \\
				** & ***\\
			\end{pmatrix}
		}_S,
		\end{align*}
		where the matrix $S$ is defined by 
		\begin{align}
		S:=\left(\begin{smallmatrix}
		\bar{H}_\mathrm{P} & 0 \\
		0 & D 
		\end{smallmatrix}\right)-\bar{H}_{\mathrm{FCI}}
		\end{align}
		The matrix block $D$ is chosen as a diagonal matrix with the otherwise irrelevant FCI eigenvalues on the diagonal in order to isolate them from the relevant eigenvalues. Hence for $\varepsilon=0$ the matrix $\bar{H}(\varepsilon)$ returns the FCI eigenvalues. The perturbation is invoked by $\varepsilon > 0$.  For $\varepsilon=1$ the matrix $\bar{H}(\varepsilon)$ provides the CCSD eigenvalues.\\
		Based on the mathematical results from section \ref{subsec: Mathematical Background}, the transition from the FCI to the CCSD eigenvalues can be characterized in more detail. 
		The FCI eigenvalues are connected to the CCSD eigenvalues in a continuous manner (see theorem \ref{theorem: locale series}). For the connection it applies that in the neighbourhood of each simple real eigenvalue only real eigenvalues occur (see theorem \ref{theorem: real neigborhood}) and that complex eigenvalues arise only if two eigenvalues coincide (see theorem \ref{theorem: multiple eigenvalues}). Here it plays an important role for the development of the eigenvalues whether the matrix is defective at this point or not (see theorem \ref{theorem: smooth path}). Altogether the following five scenarios can be sketched:
	
		\begin{figure}
			\label{fig: Schematic representation}
			\centering %l u r o
			\includegraphics[width=0.5\textwidth, trim = 10mm 10mm 40mm 150mm,clip, clip]{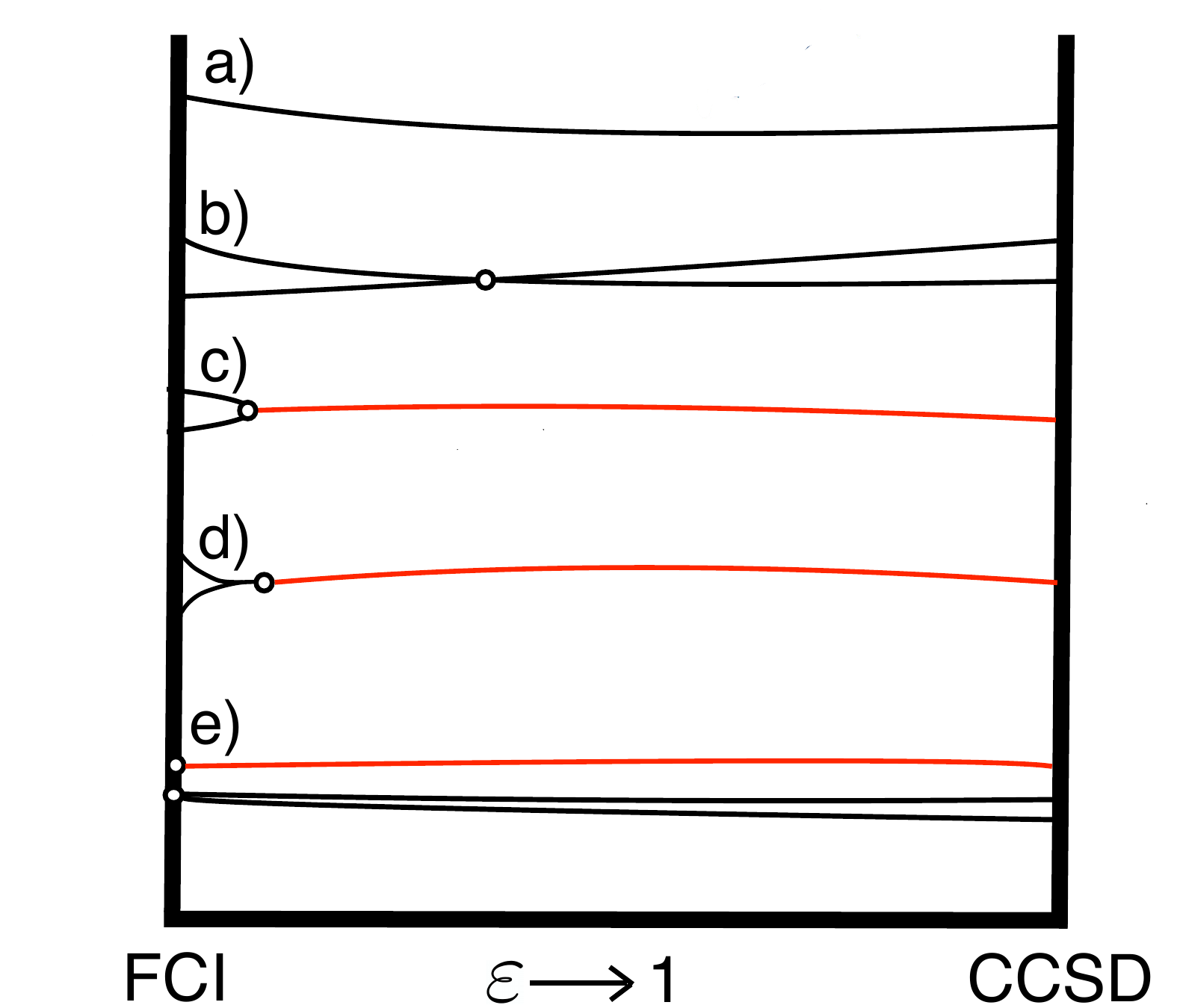}\\
			\caption{Schematic representation of the possible connections between the FCI  and CCSD energy values in the case where the matrix $\bar H(\epsilon)$ has only real entries.}
		\end{figure}

			\begin{itemize}
				\item[a)] A simple FCI eigenvalue is sufficiently well separated from the other FCI eigenvalues so that it does not coincide with any other on the connection to the CCSD eigenvalues. Then the corresponding EOM-CCSD energy value is simple and real.
				\item[b)] Two eigenvalues on the connection between FCI and CCSD eigenvalues coincide without complex eigenvalues arising in the neighbourhood of the multiple eigenvalue. Even then the energy value of the EOM-CCSD method is real as in scenario a). 
				\item[c)] 
				On the connection between FCI and CCSD eigenvalues two eigenvalues coincide for an $\varepsilon_0 \in [0,1]$ for which the matrix $\bar{H}(\varepsilon_0)$ is defective and for which in the neighbourhood of $\varepsilon_0$ complex eigenvalues occur. In contrast to scenarios a) and b), the series development of the eigenvalue $\lambda(\varepsilon_0)$ is then dominated by the term $\varepsilon^{0.5}$ (see theorem \ref{theorem: multiple eigenvalues}). This leads to large differences between CCSD and FCI eigenvalues and to complex CCSD eigenvalues.
				% note: it can happen that they become again real, if another point with defective matrices occur
				% along the connnection path
				\item[d)] Similar to scenario c) complex eigenvalues occur, but the matrix is not defective at the point where multiple eigenvalues occur. In this case, a complex part of the eigenvalue is generated at the earliest by the term $\varepsilon^\frac{3}{2}$ in the series development (see theorem \ref{theorem: smooth path}).                             
				\item[e)] The FCI method provides a multiple eigenvalue. Due to the fact that the $\bar{H}_{\mathrm{FCI}}$ matrix is non-defective (since it is a similarity transformation of the Hermitian matrix $H_\mathrm{FCI}$), the series development of $\lambda(0)$ is dominated by the term $\varepsilon^1$. Both complex or real CCSD energy values are possible. 
		\end{itemize}
	
		The connection between FCI and CCSD eigenvalues can be illustrated as in Figure \ref{fig: Schematic representation}. A black line indicates here real eigenvalues for the matrix $H(\varepsilon)$, a red line the appearance of a pair of complex-conjugated eigenvalues, and \glqq $ \circ $ \grqq {} marks the occurrence of a multiple eigenvalue.\\
		Two facts become clear from this analysis.
		First, if a FCI energy value is well separated from all others the corresponding EOM-CCSD energy value is real (see Scenario a)). This is the reason why complex energy values rarely occur in EOM-CCSD calculations. Together with the assumption that the energy gap between ground-state energy and the first excitation energy is sufficiently large it leads to the fact that the CC ground-state energy is real-valued. This is in agreement  with the statement cited in Section \ref{subsec: Coupled cluster} from Schneider's results.\cite{Schneider2009} 
		Second, if a complex EOM-CCSD energy value occurs, a square-root like behaviour results, as in scenario d), caused by the term $\varepsilon^\frac{1}{2}$ in the series expansion of the eigenvalue. This leads to large discrepancies between FCI and CCSD energy values. \\
		
		In practice, the described connection between the FCI and CCSD methods is difficult to investigate. For this reason, we introduce in the following an artificial system, which consists of four electrons and six possible states. The FCI space is thus spanned by six Slater determinants $\{\psi_0,... ,\psi_5\}$. The situation can be illustrated with a MO-like representation:\\
		\begin{minipage}{1.0\textwidth}
			\scalebox{0.65}{
				\begin{minipage}{.003\textwidth}
					%\tiny
					\addtolength{\jot}{0.5em}		
				\end{minipage}%
				\hspace*{0.5cm}
				\begin{minipage}{0.02\textwidth}
					\centering
					\vspace*{0.7cm}
					\begin{tikzpicture}	[scale=0.55]		
					\drawLevel[elec =  up, pos = {(0,3)}, width=0.6] {1sH}		
					\drawLevel[elec =  up, pos = {(0,4.5)}, width=0.6] {1sH}
					\drawLevel[elec =  no, pos = {(0,6)}, width=0.6] {1sH}
					\drawLevel[elec =  no, pos = {(0,7.5)}, width=0.6] {1sH}
					\end{tikzpicture}		
				\end{minipage}
				\begin{minipage}{0.02\textwidth}
					\centering
					\vspace*{0.7cm}
					\begin{tikzpicture}	[scale=0.55]		
					\drawLevel[elec =  down, pos = {(0,3)}, width=0.6] {1sH}		
					\drawLevel[elec =  down, pos = {(0,4.5)}, width=0.6] {1sH}
					\drawLevel[elec =  no, pos = {(0,6)}, width=0.6] {1sH}
					\drawLevel[elec =  no, pos = {(0,7.5)}, width=0.6] {1sH}
					\end{tikzpicture}
				\end{minipage}%
				\hspace*{1.0cm}
				\begin{minipage}{0.02\textwidth}
					\centering
					\vspace*{0.7cm}
					\begin{tikzpicture}	[scale=0.55]		
					\drawLevel[elec =  no, pos = {(0,3)}, width=0.6] {1sH}		
					\drawLevel[elec =  up, pos = {(0,4.5)}, width=0.6] {1sH}
					\drawLevel[elec =  up, pos = {(0,6)}, width=0.6] {1sH}
					\drawLevel[elec =  no, pos = {(0,7.5)}, width=0.6] {1sH}
					\end{tikzpicture}
				\end{minipage}
				\begin{minipage}{0.02\textwidth}
					\centering
					\vspace*{0.7cm}
					\begin{tikzpicture}	[scale=0.55]		
					\drawLevel[elec =  no, pos = {(0,3)}, width=0.6] {1sH}		
					\drawLevel[elec =  down, pos = {(0,4.5)}, width=0.6] {1sH}
					\drawLevel[elec =  down, pos = {(0,6)}, width=0.6] {1sH}
					\drawLevel[elec =  no, pos = {(0,7.5)}, width=0.6] {1sH}
					\end{tikzpicture}
				\end{minipage}
				\hspace*{1.0cm}
				\begin{minipage}{0.02\textwidth}
					\centering
					\vspace*{0.7cm}
					\begin{tikzpicture}	[scale=0.55]		
					\drawLevel[elec =  no, pos = {(0,3)}, width=0.6] {1sH}		
					\drawLevel[elec =  up, pos = {(0,4.5)}, width=0.6] {1sH}
					\drawLevel[elec =  no, pos = {(0,6)}, width=0.6] {1sH}
					\drawLevel[elec =  up, pos = {(0,7.5)}, width=0.6] {1sH}
					\end{tikzpicture}
				\end{minipage}
				\begin{minipage}{0.02\textwidth}
					\centering
					\vspace*{0.7cm}
					\begin{tikzpicture}	[scale=0.55]		
					\drawLevel[elec =  no, pos = {(0,3)}, width=0.6] {1sH}		
					\drawLevel[elec =  down, pos = {(0,4.5)}, width=0.6] {1sH}
					\drawLevel[elec =  no, pos = {(0,6)}, width=0.6] {1sH}
					\drawLevel[elec =  down, pos = {(0,7.5)}, width=0.6] {1sH}
					\end{tikzpicture}
				\end{minipage}%
				\hspace*{1.0cm} 
				\begin{minipage}{0.02\textwidth}
					\centering
					\vspace*{0.7cm}
					\begin{tikzpicture}	[scale=0.55]		
					\drawLevel[elec =  up, pos = {(0,3)}, width=0.6] {1sH}		
					\drawLevel[elec =  no, pos = {(0,4.5)}, width=0.6] {1sH}
					\drawLevel[elec =  up, pos = {(0,6)}, width=0.6] {1sH}
					\drawLevel[elec =  no, pos = {(0,7.5)}, width=0.6] {1sH}
					\end{tikzpicture}
				\end{minipage}
				\begin{minipage}{0.02\textwidth}
					\centering
					\vspace*{0.7cm}
					\begin{tikzpicture}	[scale=0.55]		
					\drawLevel[elec =  down, pos = {(0,3)}, width=0.6] {1sH}		
					\drawLevel[elec =  no, pos = {(0,4.5)}, width=0.6] {1sH}
					\drawLevel[elec =  down, pos = {(0,6)}, width=0.6] {1sH}
					\drawLevel[elec =  no, pos = {(0,7.5)}, width=0.6] {1sH}
					\end{tikzpicture}
				\end{minipage}%
				\hspace*{1.0cm} 
				\begin{minipage}{0.02\textwidth}
					\centering
					\vspace*{0.7cm}
					\begin{tikzpicture}	[scale=0.55]		
					\drawLevel[elec =  up, pos = {(0,3)}, width=0.6] {1sH}		
					\drawLevel[elec =  no, pos = {(0,4.5)}, width=0.6] {1sH}
					\drawLevel[elec =  no, pos = {(0,6)}, width=0.6] {1sH}
					\drawLevel[elec =  up, pos = {(0,7.5)}, width=0.6] {1sH}
					\end{tikzpicture}
				\end{minipage}
				\begin{minipage}{0.02\textwidth}
					\centering
					\vspace*{0.7cm}
					\begin{tikzpicture}	[scale=0.55]		
					\drawLevel[elec =  down, pos = {(0,3)}, width=0.6] {1sH}		
					\drawLevel[elec =  no, pos = {(0,4.5)}, width=0.6] {1sH}
					\drawLevel[elec =  no, pos = {(0,6)}, width=0.6] {1sH}
					\drawLevel[elec =  down, pos = {(0,7.5)}, width=0.6] {1sH}
					\end{tikzpicture}
				\end{minipage}%
				\hspace*{1.0cm} 
				\begin{minipage}{0.02\textwidth}
					\centering
					\vspace*{0.7cm}
					\begin{tikzpicture}	[scale=0.55]		
					\drawLevel[elec =  no, pos = {(0,3)}, width=0.6] {1sH}		
					\drawLevel[elec =  no, pos = {(0,4.5)}, width=0.6] {1sH}
					\drawLevel[elec =  up, pos = {(0,6)}, width=0.6] {1sH}
					\drawLevel[elec =  up, pos = {(0,7.5)}, width=0.6] {1sH}
					\end{tikzpicture}
				\end{minipage}
				\begin{minipage}{0.02\textwidth}
					\centering
					\vspace*{0.7cm}
					\begin{tikzpicture}	[scale=0.55]		
					\drawLevel[elec =  no, pos = {(0,3)}, width=0.6] {1sH}		
					\drawLevel[elec =  no, pos = {(0,4.5)}, width=0.6] {1sH}
					\drawLevel[elec =  down, pos = {(0,6)}, width=0.6] {1sH}
					\drawLevel[elec =  down, pos = {(0,7.5)}, width=0.6] {1sH}
					\end{tikzpicture}
				\end{minipage}%
			}
			\vspace*{0.1cm}\\
			\hspace*{0.5cm} $\psi_0$ \hspace*{0.6cm} $\psi_1$ \hspace*{0.7cm} $\psi_2$ \hspace*{0.6cm} $\psi_3$ \hspace*{0.6cm} $\psi_4$ \hspace*{0.6cm} $\psi_5$
		\end{minipage}
		\vspace*{0.3cm}\\
		By specifying the matrix representation $H_{\mathrm{FCI}}$ of the Hamiltonian, the system is fully described.   
		\begin{figure}
			\label{fig: Connection between}
			\centering
			\includegraphics[trim = 20mm 74mm 20mm 60mm, angle=0, clip, width=0.45\textwidth]{pictures/legend4.pdf}\\
			\includegraphics[width=0.45\textwidth]{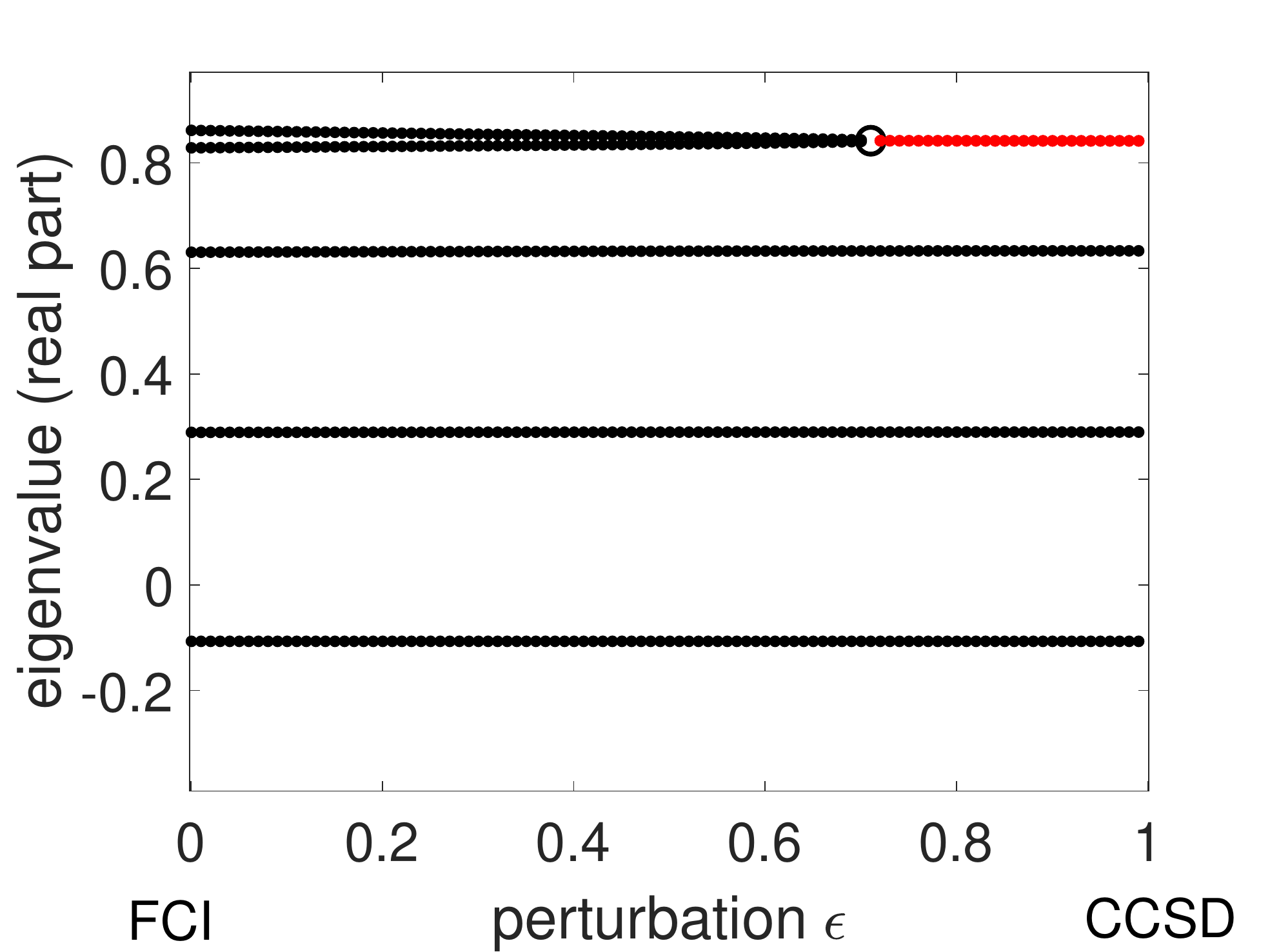}
			\includegraphics[width=0.45\textwidth]{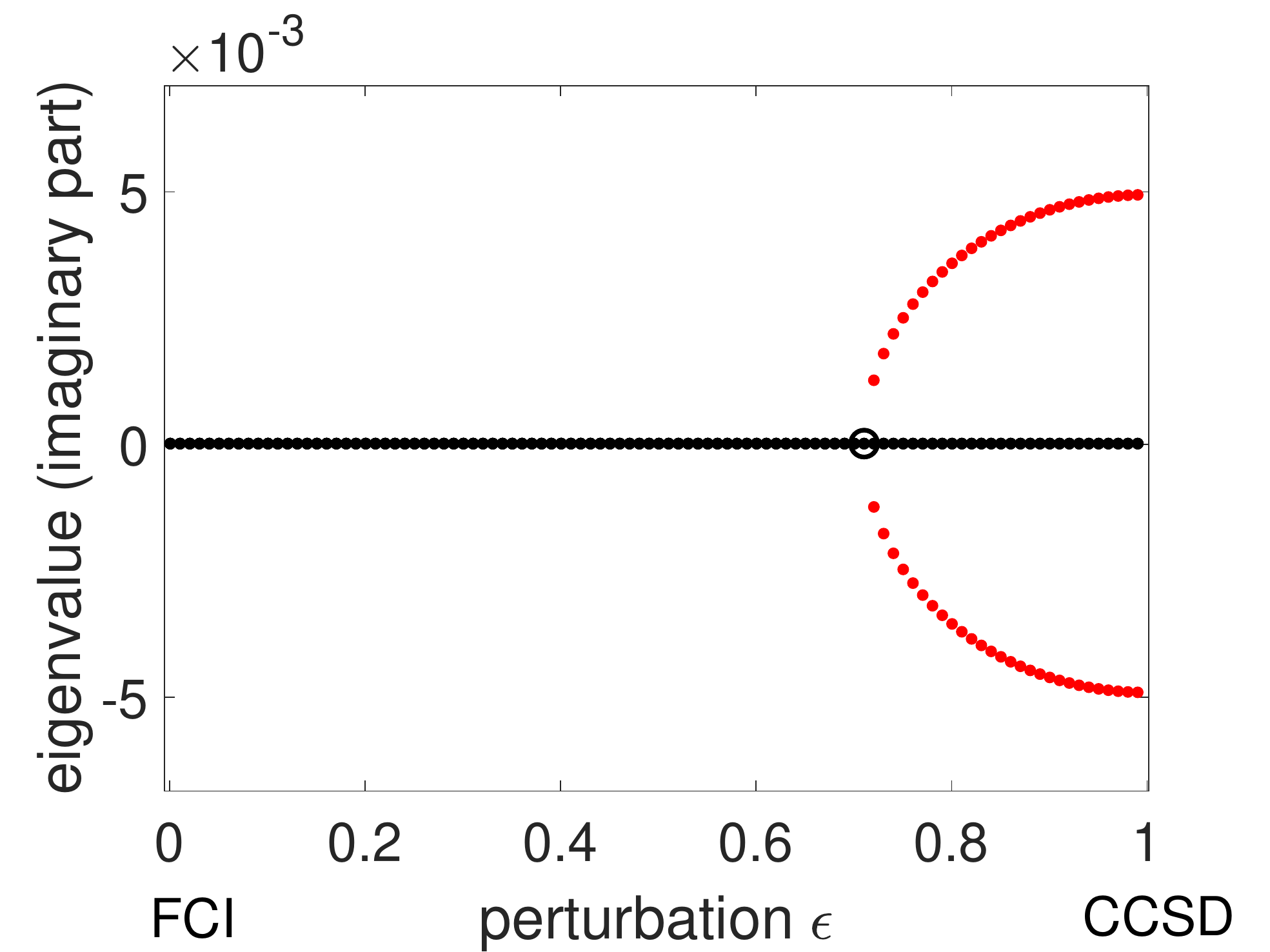}\\			\caption{Connection between FCI and CCSD eigenvalues for an sample Hamiltonian, defined by the matrix representation $H_1$ (see Appendix \ref{sec: Computational details}) for the 6-state model discussed in the text. }
		\end{figure}
		
		Using this model, we test the previously presented connection between the FCI and CCSD energy values by actual calculations. The FCI eigenvalues are obtained from the chosen $6 \times 6$ matrix representation ${H}_{\mathrm{FCI}}$ of the Hamiltonian (see Appendix \ref{sec: Computational details} for details of ${H}_{\mathrm{FCI}}$). The CCSD eigenvalues result from the truncated $5 \times 5$ matrix $\bar{H}_\mathrm{P}$ (see Section \ref{subsec: Coupled cluster}).  The amplitudes of $\hat{T}_2$ are determined via a standard CC calculation. 
		The results of the calculation are shown in Figure~\ref{fig: Connection between} and are consistent with the theoretical predictions.	
		The lowest three energy values are well-separated from all others, no complex energy values occur and a deviation between FCI and CCSD energy values is hardly recognizable. However, the two highest energy values are not well separated in the FCI solution. Here a multiple eigenvalue occurs for $\bar{H}(\varepsilon \approx 0.7)$ and for the CCSD method complex energy values arise. As predicted, a square-root-like behaviour can be observed and the difference between the energy values of the FCI and CCSD solution is clearly visible.
		
		\subsection{Complex energy values in the vicinity of conical intersections}
		\label{subsec: Complex energy}
\begin{figure} %l b r t     t l b r 
				\centering
				\label{fig: Around the}
				\includegraphics[width=0.4\textwidth,  angle=0, trim = 42mm 186mm 16mm 10mm, clip]{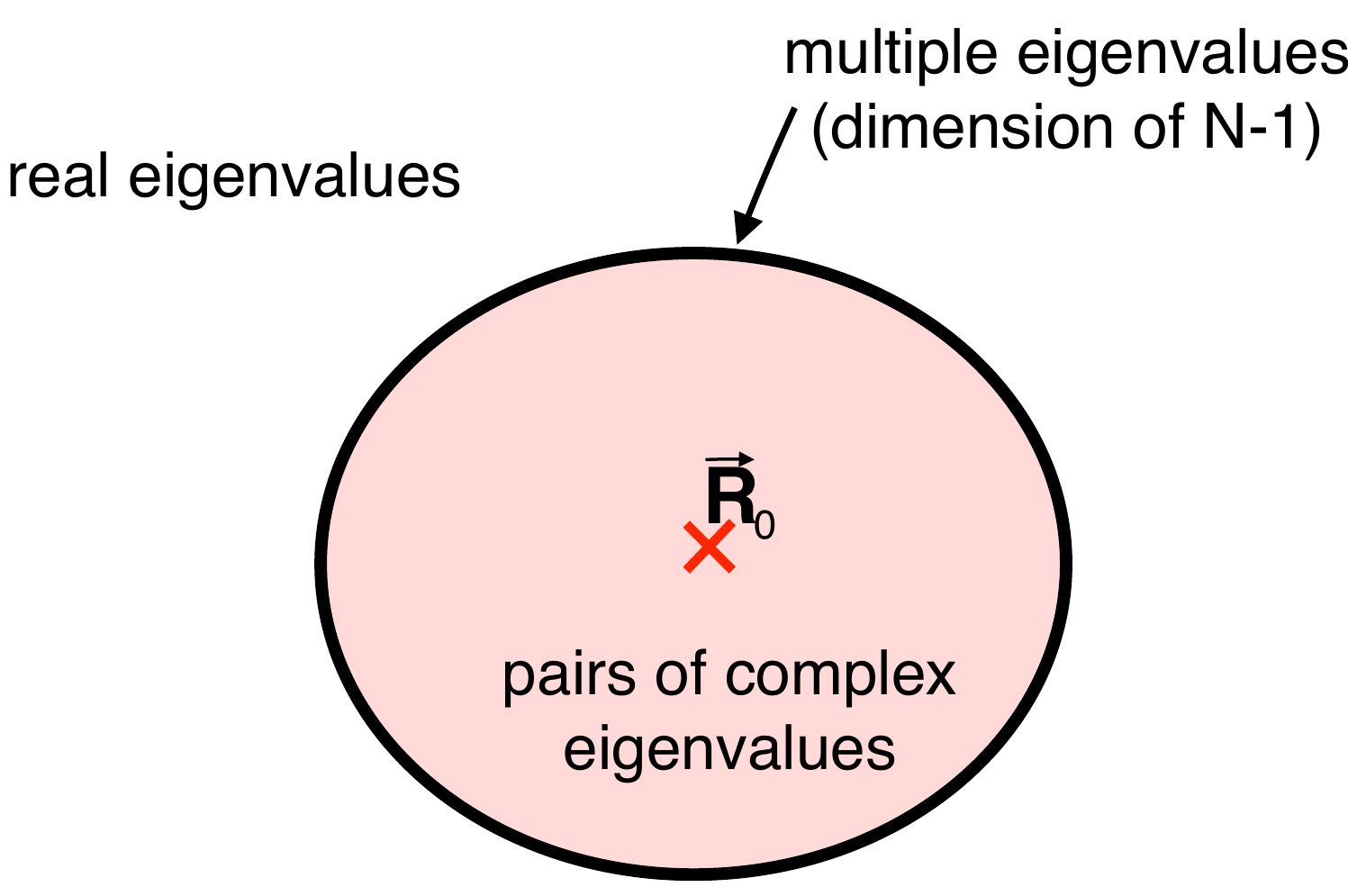}		
				\caption{In the neighbourhood of any given geometry $\vec{R}_0$, for which a complex energy value is found, pairs of complex eigenvalues occur.  Since real eigenvalues exist at some distance from $\vec{R}_0$, there is an enveloping space of dimension $N-1$, where multiple eigenvalues arise. }		
			\end{figure}

			The occurrence of a multiple eigenvalue in CC calculations implies that two different states of the molecule have close-lying energy values and that the two corresponding potential surfaces intersect. In three-dimensional space, the shape of the two potential surfaces at the point of intersection results in two cones placed on top of each other, touching each other at the tips (see Figure \ref{fig: Qualitatively right}). Therefore, the phenomenon is called conical intersection.\cite{Yarkony96, Yarkony2001, Truhlar03, Yarkony16} This is the only context in which complex energy values in CC calculations have been discussed and observed so far.\cite{Haettig2005, Koehn2007, KMMK2017} 
			Our investigation goes beyond a two-state model (see Section \ref{subsec: Mathematical Background}). 
			%In contrast to the previous literature, we do not use a two-state model, but use our mathematical results as outlined in section \ref{subsec: Mathematical Background}. 
			A short comparison to the two-state model can be found in Section \ref{sec: comparison to two state model}.
			First, let us examine why complex energy values can occur exclusively in the context of conical intersections. We assume %by assuming 
			that the matrix entries of the Hamiltonian matrix depend in a continuous manner on the molecular geometry:\\ 
			Suppose there exists a molecular geometry $ \vec {R} _0 $ for which a complex eigenvalue $ E_k $ of the real $\bar{H}_\mathrm{P}$ matrix occurs.
			As the matrix entries depend continuously on molecular geometrical parameters, there exists a neighbourhood of $ \vec {R} _0 $ where the eigenvalue $E_k$ maintains a non-vanishing imaginary part (see Figure \ref{fig: Around the}). 
			In this neighbourhood, the complex-conjugate eigenvalue also occurs (see Theorem \ref{theorem: real neigborhood}).
			To enter the domain of real eigenvalues, at least two eigenvalues need to coincide (see Theorem \ref{theorem: multiple eigenvalues}). For this reason, around the point $\vec{R}_0$ an enveloping space of geometries with multiple eigenvalues exists, which can be seen as a set of intersection points, commonly called intersection seam. \\
			By continuing this discussion, the following conclusion can be drawn about the dimension of the intersection seam.  Let $N$ denote the number of geometrical degrees of freedom of the molecule.
			Since the enveloping-space surrounds the geometry $\vec{R}_0$ one can conclude that the dimension of the enveloping space is $N-1$.
			However, quantum mechanics as well Hermitian quantum-chemical theories require that the intersection seam has a dimension of $N-2$.\cite{Neumann29,Yarkony96,Keller2008} Therefore a complex eigenvalue of a real $\bar{H}_\mathrm{P}$ matrix leads to a qualitatively wrong representation of the potential surface. \\
			
		\subsection{Shape of conical intersections}
		\label{subsec: Shape of}
		For the shape of conical intersections the following is known so far. On the one hand, if the matrix $\bar{H}_\mathrm{P}$ is non-defective at the intersection point, the energy gap between the two states in the neighbourhood of the intersection point is linear with the distance from the intersection point\cite{KMMK2017} as required by quantum mechanics.\cite{Teller37}
		On the other hand, if the $\bar{H}_\mathrm{P}$ matrix is defective, numerical investigations\cite{Koehn2007, KMMK2017, KK2017} show a root-like behaviour of the potential surface. In previous investigations both aspects were considered separately and the root-like behaviour was observed but not rationalized by means of a detailed theoretical analysis. With the results from Section~\ref{sec: Theory} both aspects can be derived mathematically at the same time:\\	
			
		Let us assume that the entries of $\bar{H}_\mathrm{P}$ depend analytically on the geometrical parameters of the molecule, since the Hamiltonian depends analytically on the internal coordinates. Varying a given geometry $\vec{R}_1$ in one direction thus can be described by a real parameter $r$. The corresponding energy values are determined by the eigenvalues $\lambda_i(r)$ of the matrix $\bar{H}_\mathrm{P}(r)$. 
		Let us assume furthermore that at a geometry $\vec{R}_1$ an intersection  occurs, which means that $\lambda(0)$ is a multiple eigenvalue of $\bar{H}_\mathrm{P}(0)$. The shape of the conical intersection at $\vec{R}_1$ is determined by the eigenvalues $\lambda_i(r)$, and depends on the properties of the matrix $\bar{H}_\mathrm{P}(0)$: 
		\begin{itemize}
			\item In case of a defective matrix $ \bar{H}_\mathrm{P}(0) $, according to Theorem \ref{theorem: locale series}, the series expansion of the eigenvalue $\lambda(r)$ starts in certain cases with a term proportional to $r^{\frac{1}{2}}$. This leads to a square-root-like
			evolution of the potential surface close to the intersection seam %behaviour of the potential surface 
			and to complex eigenvalues near the intersection point.\\
			\item In case of a non-defective matrix $\bar{H}_\mathrm{P}(0)$, according to Theorem \ref{theorem: smooth path}, the series expansion of the eigenvalue $\lambda(r)$ near the intersection point starts with a term proportional to $r$.  
			By subtracting the expansions of two eigenvalues
			\begin{align*}
				\lambda_1(r)=\lambda(0)+a_{1}r + a_{2}r^{1+\frac{1}{\tilde{m}_1}}+\cdots,\\
				\lambda_2(r)=\lambda(0)+b_{1}r + b_{2}r^{1+\frac{1}{\tilde{m}_2}}+\cdots
			\end{align*}
			with $\tilde{m}_1, \tilde{m}_2 \in \mathbb{N}$, $\tilde{m}_1 \ge 1, \tilde{m}_2 \ge 1$,
			the linearity of the energy gap is obvious: 
			\begin{align*}
				\lambda_1(r)-\lambda_2(r)=(a_{1}-b_{1})r+\cdots.
			\end{align*}

		\end{itemize}

		\begin{figure}	
				\label{fig: Qualitatively right}
				\centering
				\subfigure[Qualitatively correct shape of a conical intersection, as expected from an EOM-CCSD calculation with a non-defective matrix $\bar{H}_\mathrm{P}$.]{\includegraphics[trim = 25mm 65mm 16mm 49mm,angle=270, clip, width=0.40\textwidth]{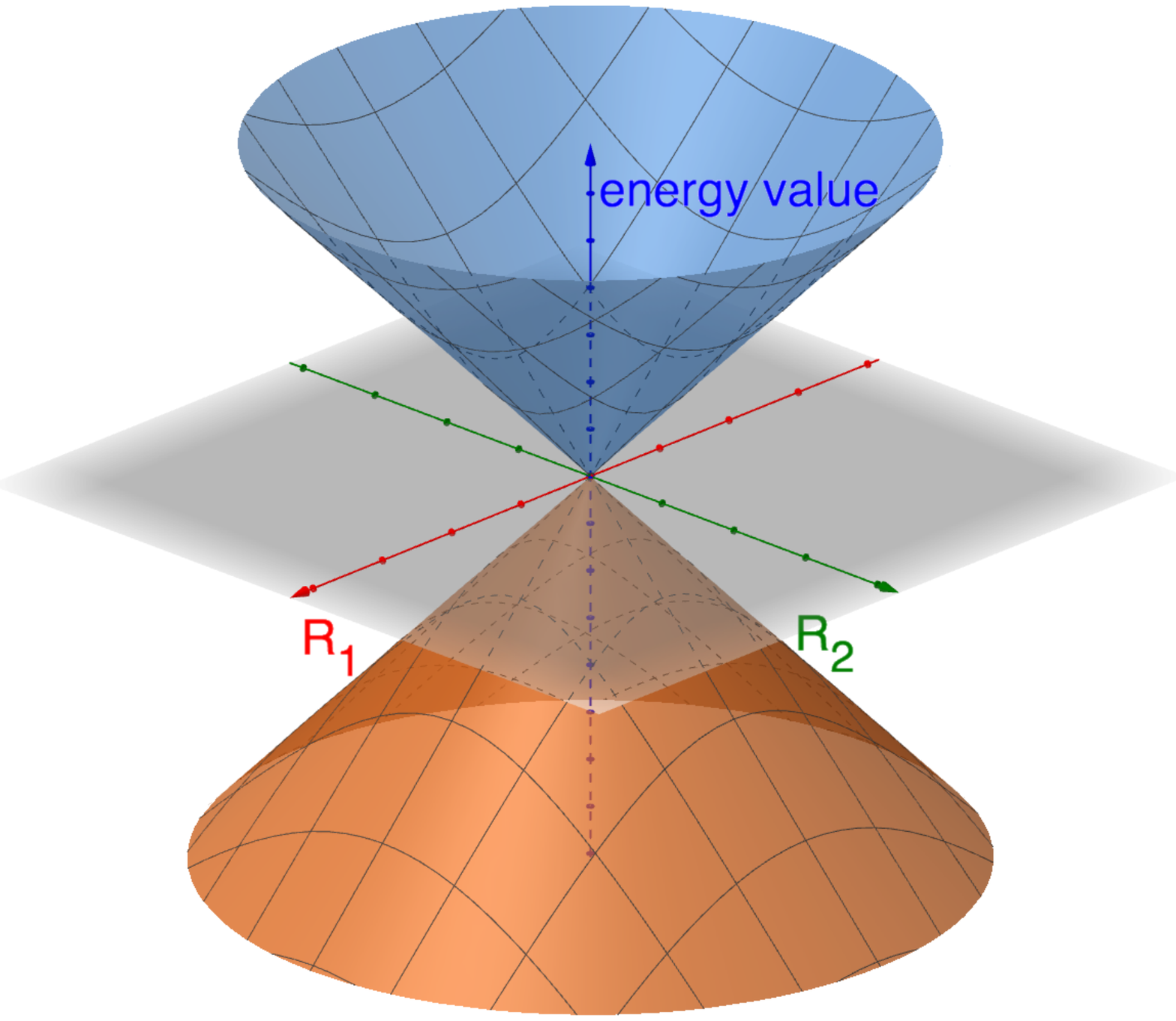}}	
				\subfigure[Qualitatively wrong shape of a conical intersection, as it may occur in an EOM-CCSD calculation with a defective matrix $\bar{H}_\mathrm{P}$.]{\includegraphics[trim = 25mm 65mm 16mm 49mm, , clip, width=0.35\textwidth]{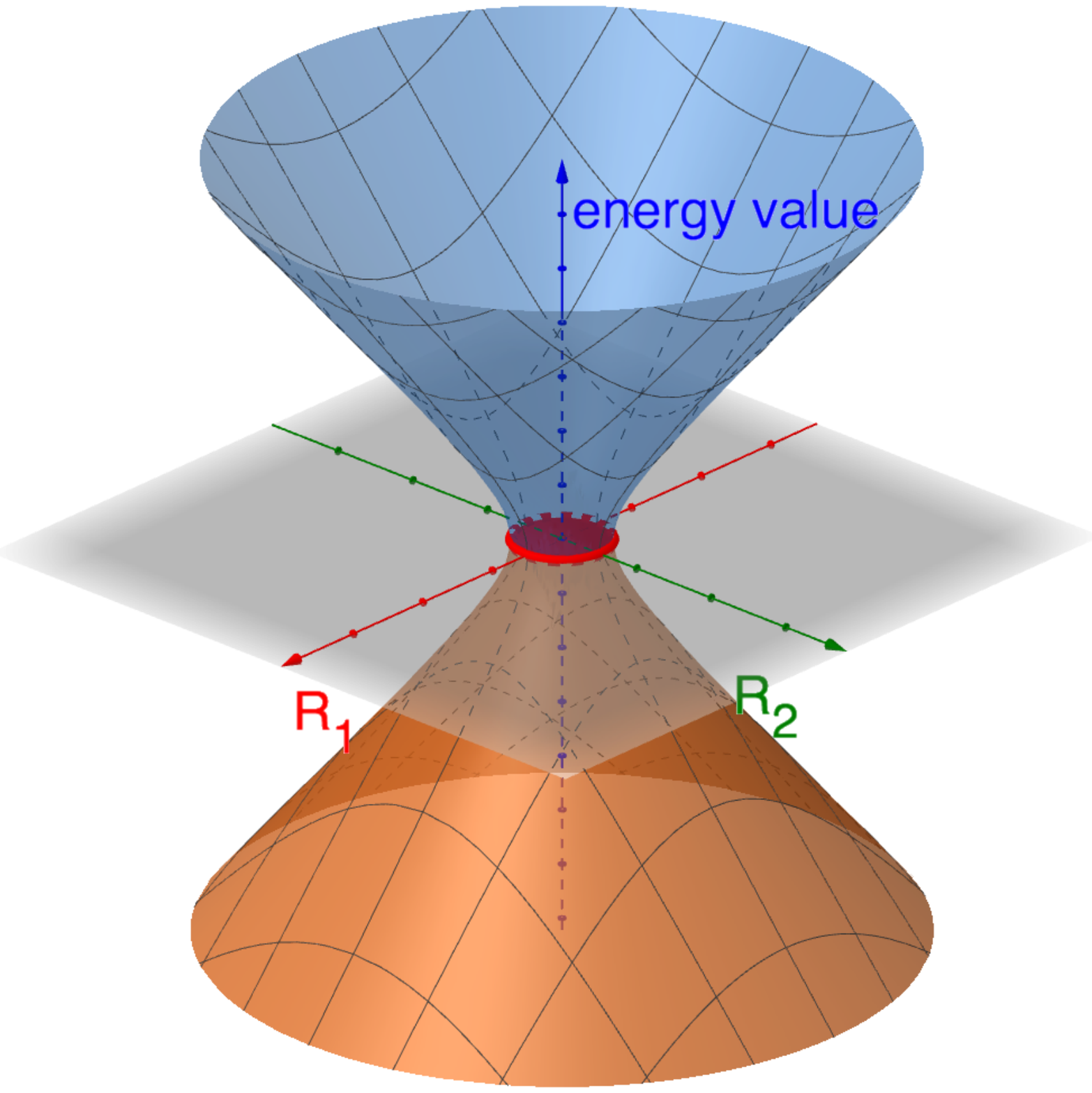}}
				\caption{Comparison between the shape of a conical intersection in case of a defective and a non-defective matrix $\bar{H}_\mathrm{P}$ for a system with two degrees of freedom 
				$R_1 $ and $ R_2 $.}	
		\end{figure} 
		Figure \ref{fig: Qualitatively right} illustrates the expected behaviour of the potential surfaces in the case of two degrees of freedom for a defective and non-defective matrix. 
		Case $(a)$ shows the correct shape of the potential surface. The dimension of the intersection seam is zero ($N-2$). The linear behaviour close to the intersection point is also clearly visible.
		A possible shape for the case of a defective matrix is shown in \ref{fig: Qualitatively right} $(b)$. Within the red-coloured disk, pairs of complex-conjugated eigenvalues occur. In the enveloping space, located at the edge of the disk multiple eigenvalues occur. Looking at a cut through this figure, the root-like behaviour along each direction becomes clear. Here all $\bar{H}_\mathrm{P}$ matrices based on the molecular geometries located at the edge of the disc are defective.\\
		
        At this point the question arises in which cases the matrix becomes defective. 
        For the case that the eigenvectors describe crossing states of different symmetries, they cannot become linearly dependent. Assuming that none of the other eigenvalues coincide at this point the matrix is non-defective.
        In the absence of a constraint that ensures linear independence of the eigenvectors, the defectiveness of the matrix seems to be rather the rule than the exception for the following reasons:
        First, in general, to ensure a non-defective matrix with multiple eigenvalues more conditions have to be fulfilled than for a defective one.\cite{Keller2008}
        Second, in numerical examples providing crossing states of the same symmetry defective matrices occur in all cases.\cite{Koehn2007, KMMK2017}\\
        
        % was heißt symmetrical methods, references for adc and unitary CC
        On one hand, to avoid defectiveness of the similarity transformed Hamiltonian matrix, symmetric methods (i.e., those derived in the algebraic diagrammatic construction\cite{adc2} or unitary CC contexts\cite{ucc4,ucc,ucclaura}) can be used. On the other hand, a recent paper\cite{KK2017} suggests to
        ensure non-defectiveness by including additional constraints in the CC equations.

		\subsection{Comparison to the two-state model}
		\label{sec: comparison to two state model}
		In the literature,\cite{Haettig2005, Koehn2007, KMMK2017} the behaviour of CC methods at conical intersections has been analyzed using a two-state model. The behaviour in the vicinity of conical intersections is there discussed using the $2 \times 2$ matrix $H^{2\times2}(\vec{R})$ with $\vec{R}$ specifying the molecular geometry near the intersection point $\vec{R}_0$. From a mathematical point of view this is justified because the solution of \begin{equation}
		    H^{2\times2}(\vec{R}) \vec{v}(\vec{R}) = \lambda(\vec{R}) \vec{v} (\vec{R})
		\end{equation} provides under certain conditions a first-order approximation of the entire eigenvalue problem \begin{equation}
		    \bar{H}_P(\vec{R})\vec{v}(\vec{R})=\lambda(\vec{R}) \vec{v}(\vec{R}).
		\end{equation} Assuming that the entries of $\bar{H}_P(\vec{R})$ depend analytically on $\vec{R}$ and that $\bar{H}_P(\vec{R}_0)$ is non-defective this can be proven analogously to the CI case.\cite{Yarkony16}
 		In the two-state model, $H^{2\times2}(\vec{R})$ has the form
		\begin{align}
		H^{2\times2}(\vec{R}) = \begin{pmatrix}
		H_{11}(\vec{R}) & H_{12}(\vec{R}) \\
		H_{21}(\vec{R}) & H_{22}(\vec{R})\\
		\end{pmatrix}.
		\end{align}
		The eigenvalues for a given geometry $\vec{R}$ are
		\begin{align}
		    \lambda_{1,2}=\frac{1}{2}(H_{11}+H_{22}) \pm X,
		\end{align}
		    with
		\begin{align}
		    X = \frac{1}{2}\sqrt{(H_{11} - H_{22})^2+4H_{12}H_{21}}.
		\end{align}
		By means of a detailed analysis of the eigenvalues and eigenvectors of the matrix ${H}^{2\times2}$ several conclusions have been drawn.\cite{Haettig2005, Koehn2007, KMMK2017} For these conclusions, it should be noted that the model is simply a first-order approximation to the problem and that higher-order %contributions could affect them. 
		effects are not captured. 
		Based on the two-state model the crossing conditions of a conical intersection in a CC calculation were derived leading to the conclusion that for a non-defective  ${H}^{2\times2}$ matrix the dimension of the intersection seam is $N-2$ even if the matrix is not symmetric. This result was not %achieved by discussing the entire matrix as in the previous section, but might be anticipated by for the entire matrix based on the existing mathematical literature. \cite{Keller2008}
		obtained in our analysis of the entire matrix. Based on the existing mathematical literature,\cite{Keller2008} it is however likely that it holds for the entire matrix as well. 
		Both by analysis using 
		%By means of 
		the two-state model\cite{KMMK2017} as well as by analyzing the entire matrix (as done in the previous section) an energy gap near the intersection point is found that linearly depends on the distance to the intersection. %can be stated. 
		In addition, our analysis of the entire matrix 
		%our discussing for the entire matrix 
		explains the square-root-like behavior of the potential surface in the case of a defective matrix. The latter is not possible by using the two-state model alone since the model does not actually hold for the case of a defective matrix.\cite{Yarkony16} 
		
	\section{Complex values in case of complex-valued $\bar{H}_\mathrm{P}$ matrix}
\label{sec::complexValuesComplHam}
We now turn to the case for which the matrix representation $\bar{H}_\mathrm{FCI}$ of the Hamiltonian in the FCI space has complex-valued entries.
The decisive difference for the eigenvalues of a complex matrix in contrast to a matrix with only real entries is that complex eigenvalues do not have to occur in pairs. 
As a consequence of non-pairwise complex eigenvalues a real eigenvalue can gain an imaginary part through an analytic perturbation of the matrix entries without a multiple eigenvalue occurring in between. This is not possible for matrices with only real entries since then Theorem \ref{theorem: real neigborhood} applies. This leads to the following important statement:\\
In case of a complex Hamiltonian matrix $\bar{H}_\mathrm{P}$ for each energy value (even ground-state energies) of the CC methods an imaginary part may occur, even if the state level is well isolated from other states.\\

\begin{figure}
	\label{fig: Connection between 2}
	\centering
	\includegraphics[trim = 20mm 119mm 10mm 24mm, angle=0, clip, width=0.5\textwidth]{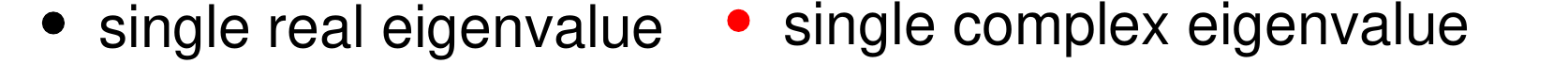}\\
	\includegraphics[width=0.45\textwidth]{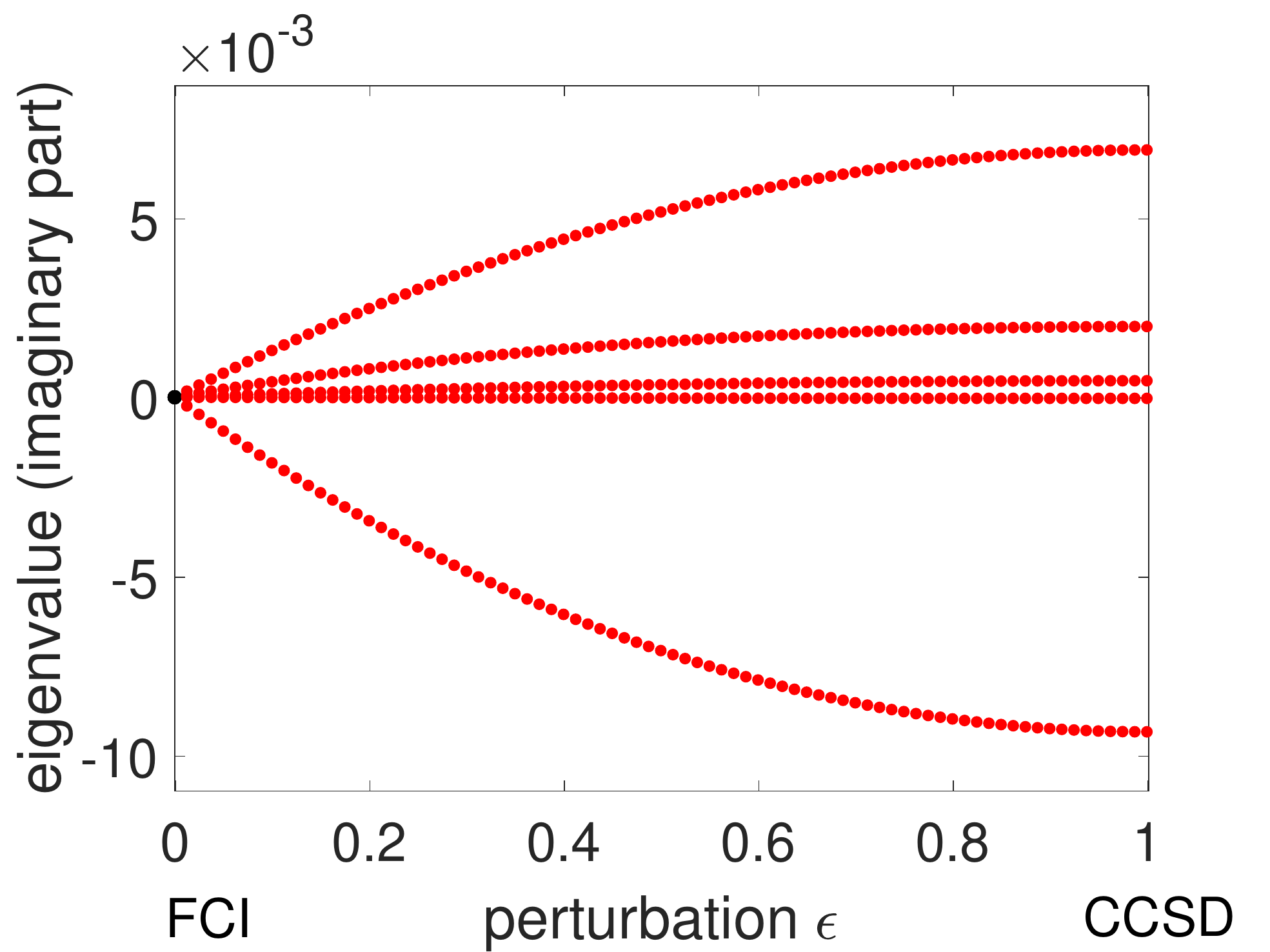}\\
	\includegraphics[width=0.45\textwidth]{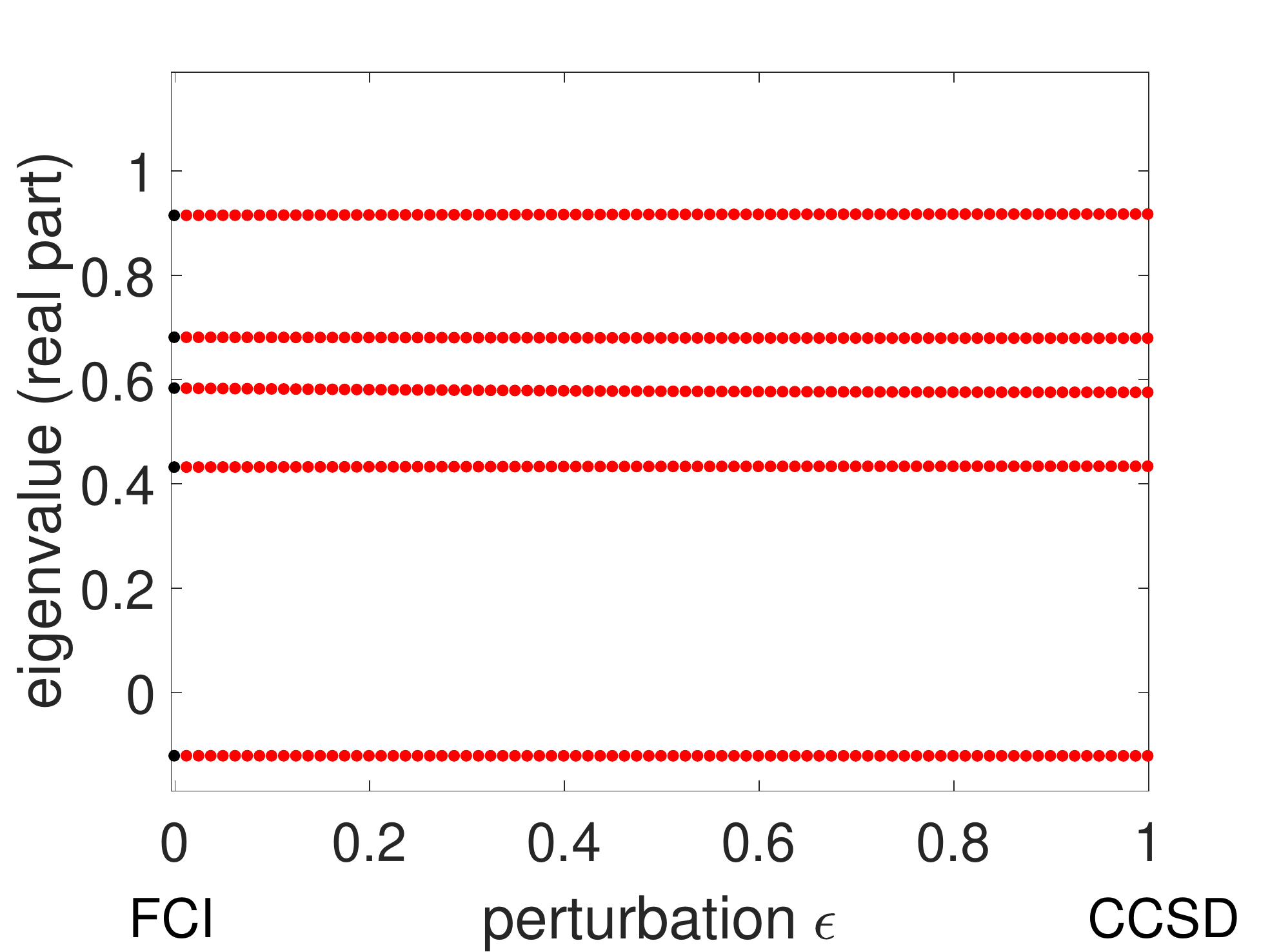}\\				\caption{Connection between FCI and CCSD eigenvalues for a complex-valued sample matrix $\bar{H}_2$ of the 6-state model.}
\end{figure}
\begin{figure}
	\label{fig: detail view}
	\centering
	\includegraphics[trim = 25mm 83mm 30mm 60mm, angle=0, clip, width=0.45\textwidth]{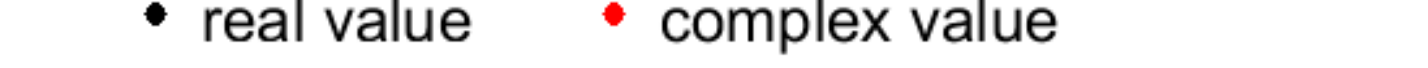}\\
	\includegraphics[width=0.45\textwidth]{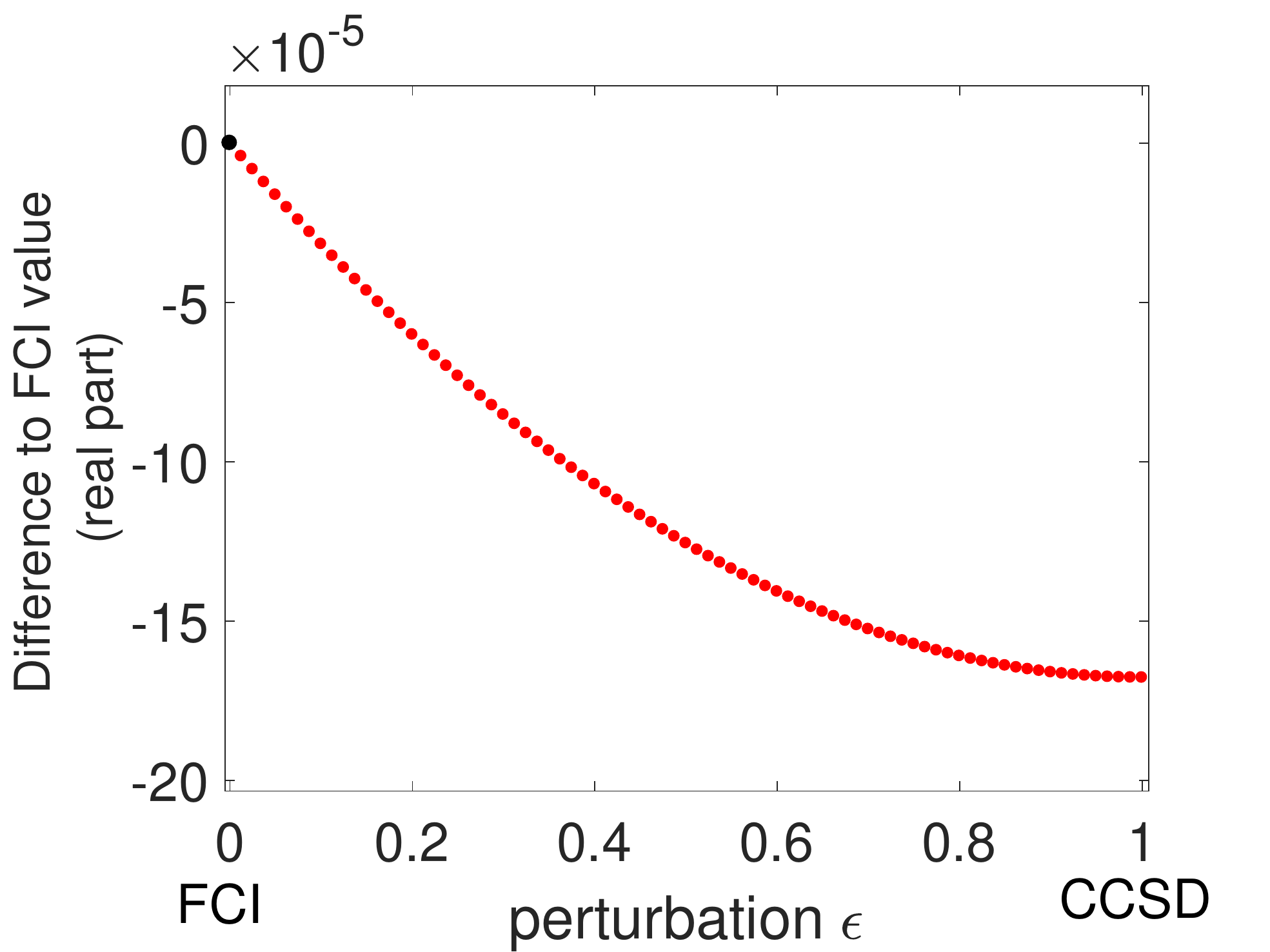}
	\includegraphics[width=0.45\textwidth]{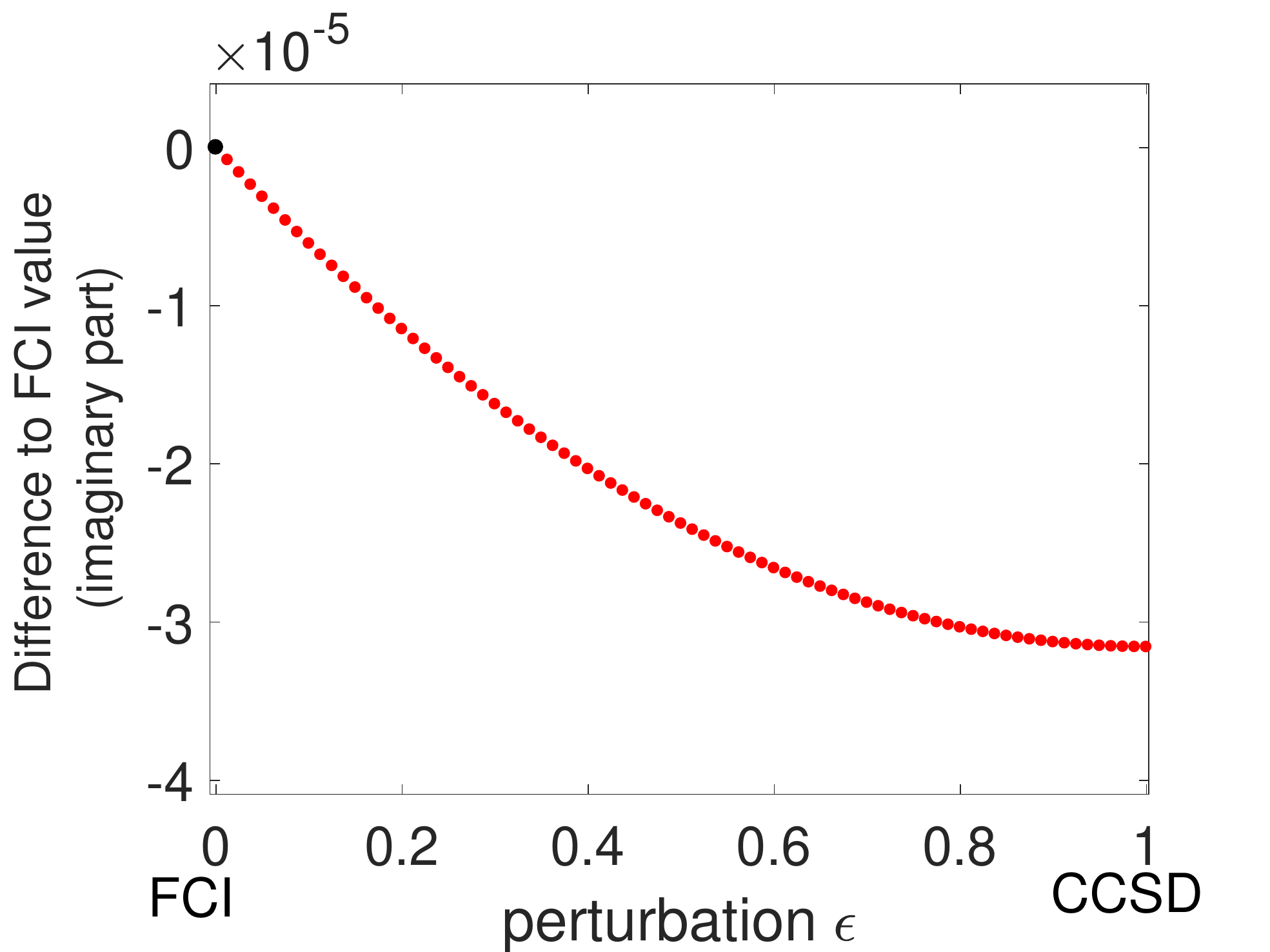}
	\caption{Development of the difference between FCI and CCSD ground-state energy values for the 6-state model. The resulting error behaves in a similar way for both the real and the imaginary parts of the energy differences. }
\end{figure}

The connection between FCI eigenvalues and CCSD eigenvalues established in Section \ref{subsec: Connection between} also applies to matrices with complex matrix entries. 
A sample calculation with a given complex-valued Hamilton matrix $H_\mathrm{FCI}$ (see Appendix \ref{sec: Computational details}) illustrates the continuous transition from FCI eigenvalues to CCSD eigenvalues for the 6-state model (see Section \ref{subsec: Connection between}, see Figure \ref{fig: Connection between 2}).
Here only the FCI eigenvalues are real. As soon as $\varepsilon$ is non-zero, complex eigenvalues occur. The CCSD eigenvalues are also complex.\\
The imaginary part of the CCSD eigenvalues is in all cases approximately of the same order of magnitude as the difference between the real parts of the FCI and CCSD eigenvalues (see Table \ref{tab: table2}).
The behaviour of the imaginary part of the eigenvalues of $H_\mathrm{P}(\varepsilon)$ is very similar to the development of the deviation of the real parts from the FCI reference (see Figure \ref{fig: detail view}).
Analogous to the always existing deviation between the real part of FCI eigenvalues and CCSD eigenvalues, the occurring imaginary part can be considered a kind of "numerical inaccuracy" of the EOM-CCSD method. Therefore, if the EOM-CC method provides a good approximation to the FCI values and the imaginary part is sufficiently small, a meaningful energy value can be defined via the real part of the corresponding CCSD eigenvalue. 

\subsection{Complex energy values for molecules in a strong magnetic field}
\label{subsec: Complex energy values for molecules in strong magnetic fields}
In the context of CC theory, complex entries in the Hamiltonian matrix $\bar{H}_\mathrm{P}$ arise for example due to the presence of a magnetic field\cite{S15, HS17} as well as in relativistic quantum-chemical calculations considering spin-orbit coupling.\cite{CowanGriffin66, SaueViss03, Berning00} In the case of a magnetic field the corresponding literature\cite{S15, HS17, HS19, HGS20} has so far not reported complex energy values. 
Nevertheless, complex energy values can indeed occur in finite magnetic-field calculations. We verify this by CCSD calculations for the ground-state of the $\mathrm{H_2O}$ molecule in a strong magnetic field. 
\begin{figure}
	\centering
	\label{flosPictures}
	\subfigure[Orientation of the magnetic field vector $\vec{B}$ in the water molecule.]{\includegraphics[trim = 93mm 76mm 48mm 133mm,clip, angle=270, width=0.35\textwidth]{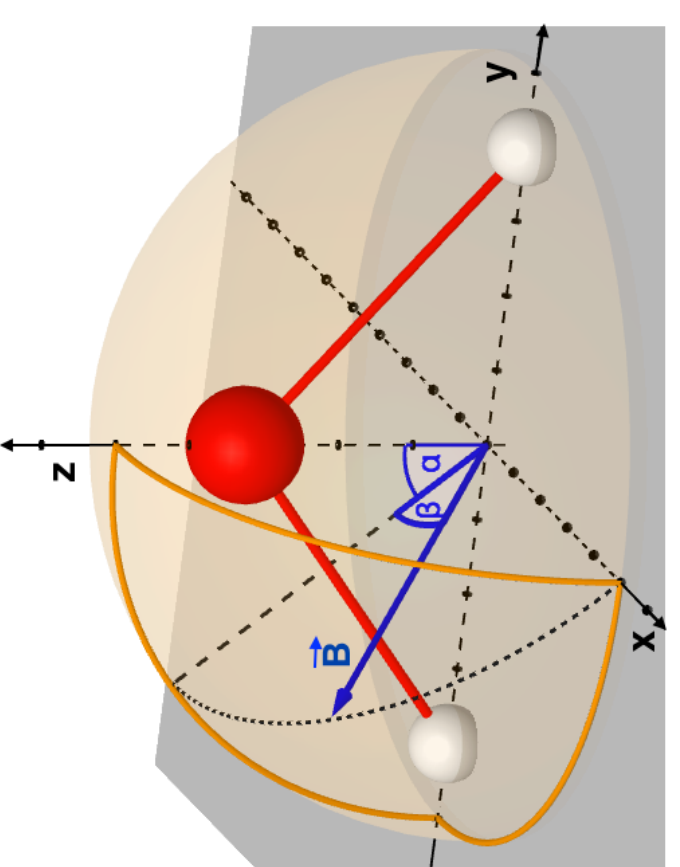}}
	\subfigure[Imaginary part of ground-state energy as results of a CCSD calculation for a water molecule in a strong magnetic field.]{\includegraphics[width=0.495\textwidth]{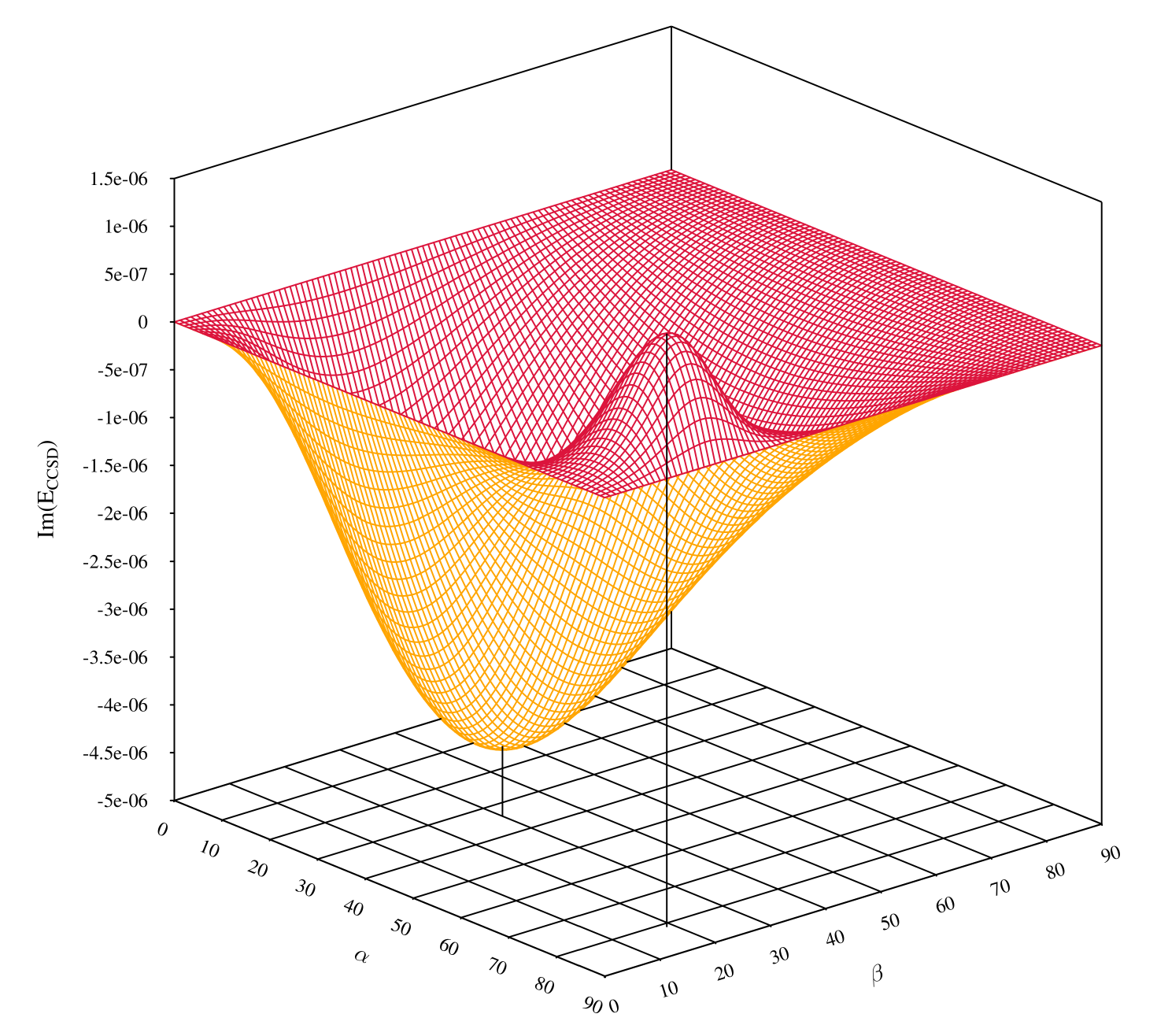}}
	\caption{Results of CCSD calculations for a water molecule in a strong magnetic field of $0.5$ $\mathrm{B_0}$ as a function of the orientation of the magnet-field vector $\vec{B}$.}
\end{figure}
For the calculations the water molecule is placed in $yz$ plane with the $C_2$ axis chosen as the $z$ direction. The alignment of the magnetic field vector $\vec{B}$ is controlled by varying the angles $\alpha$ and $\beta$ from $0^\circ$ to $90^\circ$ (see Figure \ref{flosPictures} $(a)$):
\begin{equation*}
	\vec{B}(\alpha,\beta) = 0.5 \begin{pmatrix}
		\sin(\beta)\\
		-\sin(\alpha)\cos(\beta)\\
		\cos(\alpha)\cos(\beta)\\
	\end{pmatrix}.
\end{equation*} 
The bonding angle for the water molecule is chosen to $102.21^\circ$ and the bond length to $0.9644$ $\mathrm{\mathring{A}}$.
We used a magnetic field of $0.5$ $\mathrm{B_0}$ (in this case magnetic and Coulomb forces are of similar magnitude), an uncontracted aug-cc-pVTZ Cartesian basis set \cite{Kendall92} and gauge-including atomic orbitals.\cite{London37, Tellgren12} The calculations were performed with the {\sc QCumbre} code\cite{HS17,qcumbre} for the CC part interfaced to the program package {\sc London}\cite{londonpap,london} for integral evaluation and the HF part.\\
The results of the calculation are shown in figure \ref{flosPictures}. 
The largest absolute value for the imaginary part of the energy value occurs for angles $\alpha \approx 33^\circ$ and $\beta \approx 30^\circ$ and amounts to  $4\cdot10^{-6} \mathrm{Hartree}$.  The correlation energy in this case is $3\cdot10^{-1} \mathrm{Hartree}$. This value is about 100.000 times larger than the corresponding imaginary part.
Moreover, it is noted that the obtained energy value is real if at least one component of the magnetic-field vector $\vec{B}$ vanishes. 
This observation is also made in earlier numerical studies.\cite{S15, HS17, HS19, HGS20} The fact that in these cases real energy values are obtained although the matrix has complex entries suggests that for all these cases a transformation to a real $\bar{H}_P$ matrix exists based on symmetry arguments.\\

\textbf{Symmetry-inspired transformation to a real $\bar{H}_\mathrm{P}$ matrix}\\
We now provide such a symmetry-inspired transformation that yields a real $\bar{H}_\mathrm{P}$ matrix for the H$_2$O molecule following ideas first presented by Pitzer and Winter.\cite{Pitzer87}\\
Let the water molecule with $C_{2v}$ symmetry be placed in the $yz$ plane as in our calculations at the beginning of this section (see Figure \ref{flosPictures} (a)). Let us furthermore assume that the CC calculations are performed with real-valued, symmetry-adapted basis functions $\chi$. 
Let $\boldsymbol{\chi}_{S}$ denote the set of symmetry-adapted basis functions of an arbitrary point-group symmetry $S$, thereby ignoring the applied finite magnetic field (i.e., the point group used for water is then $C_{2v}$).
%\textcolor{red}{Anmerkung Stella: Hier hab ich Diskussionsbedarf, sollte man hier noch weitere Anmerkungen machen, da ja die v-Spiegelung im Feld keine Symmetrieoperation ist: Antwort: das interessiert hier nicht, da wir nur die Irreps anschauen für die Punktgruppe ohne Berücksichtigung von B} 
The following unitary transformation will necessarily lead to a real representation of $H_\mathrm{P}$: the set of the transformed basis functions $\tilde{\boldsymbol{\chi}}_{S} $ is obtained by taking either the basis functions $\boldsymbol{\chi}_{S}$ in their original form or by multiplying the basis functions of $\boldsymbol{\chi}_{S}$ with $i$. Detailed instruction which of the sets of the basis functions need to be multiplied by $i$ are given in Table \ref{tab:transformation}. \\

Let $\tilde{\chi}_\nu$, $\tilde{\chi}_\sigma$, $\tilde{\chi}_\mu$ and $\tilde{\chi}_\rho$ be the transformed basis functions. Let us now show that with the transformation described above all relevant integrals are real:

\begin{enumerate}
	\item 	The one-electron integrals $\bra{ \tilde{\chi}_\nu}\hat{h}\ket{{ \tilde{\chi}_\mu}}$, where $\hat{h}$ is the one-electron operator including the kinetic-energy operator, the electron-nucleus repulsion, and the diamagnetic contribution due to the magnetic field. \\	
	This integral does not vanish if and only if $\tilde{\chi}_\mu$ and $\tilde{\chi}_\nu$ are part of the irreducible representation of the point group $S$. For these cases the integral is always real.
	\item The one-electron integrals
	\begin{equation}
		\langle \tilde{\chi}_{\mu} |B_x\hat{l}_x + B_y\hat{l}_y + B_z\hat{l}_z |\tilde{\chi}_{\nu}\rangle.
		\end{equation}
	Here it is important to assume that one component of the magnetic field vanishes.  As an example we consider the case $B_x=0$. The other cases can be treated in an analogous manner.  
	For the not vanishing integrals $\bra{ \tilde{\chi}_\nu} B_y \hat{l}_y\ket{{ \tilde{\chi}_\mu}}$ and $
	\bra{ \tilde{\chi}_\nu} B_z \hat{l}_z\ket{{ \tilde{\chi}_\mu}} $, exploitation of the symmetry relations leads to the following finding: 
	The integrals do not vanish if and only if $\tilde{\chi}_\mu$ is a real and $\tilde{\chi}_\nu$ is an purely imaginary basis function or vice versa.	Since $\hat{l}$ includes a $i$, the integrals are real.
	\item 
	The two-electron integrals 
	\begin{align}
	\label{equation: twoElectronIntegrals}
	\bra{ \tilde{\chi}_\nu \tilde{\chi}_\sigma}\ket{{ \tilde{\chi}_\mu \tilde{\chi}_\rho}}.
	\end{align} Here the combinations of the basic functions from the various irreducible representations must be checked. It turns out that for each possibility either the integral vanishes due to symmetry relations or the integral is real since 0, 2, or 4 of the basis functions $\tilde{\chi}_\nu$, $\tilde{\chi}_\sigma$, $\tilde{\chi}_\mu$ and $\tilde{\chi}_\rho$ are imaginary.

\end{enumerate}
The circumstance that all three different types of integrals are real leads to the following conclusions: First, the Fock matrix has only real entries and the molecular orbital coefficients during the HF calculation are real. Second, the two-electron integrals in the molecular-orbital representation are real and the matrix representation $H_\mathrm{FCI}$ is thus real as well. Third, a CC calculation provides real amplitudes (see Section \ref{subsec: Coupled cluster}). Altogether it can be stated that the $\bar{H}_\mathrm{P}$ matrix, which is based on the transformed basis set, has only real entries. Since the eigenvalues remain unchanged under the discussed basis transformation, the energy values even in case of a calculation without application of the basis transformation are real. 

\begin{table}
	\caption{Transformation of the symmetry-adapted basis functions $\chi$ for water (molecular point group is $C_{2v}$) that leads to a real $H_\mathrm{P}$ matrix in case that one component of the magnetic-field vector vanishes.}
	\label{tab:transformation}
	\begin{tabular}{l|c|c|c|c|c}
		& $\tilde{\boldsymbol{\chi}}_{A_1} $ & $\tilde{\boldsymbol{\chi}}_{A_2} $ & $\tilde{\boldsymbol{\chi}}_{B_1}$ & $\tilde{\boldsymbol{\chi}}_{B_2} $ \\
		\hline
		$B_x=0$ & $ \boldsymbol{\chi}_{A_1}$ & $ i \cdot \boldsymbol{\chi}_{A_2} $ & $ i \cdot \boldsymbol{\chi}_{B_1}  $ & $\boldsymbol{\chi}_{B_2} $  \\
		\hline
		$B_y=0$ & $\boldsymbol{\chi}_{A_1}$ & $i \cdot \boldsymbol{\chi}_{A_2}$ & $\boldsymbol{\chi}_{B_1} $ & $ i \cdot \boldsymbol{\chi}_{B_2}$ \\
		\hline
		$B_z=0$  & $\boldsymbol{\chi}_{A_1}$ & $\boldsymbol{\chi}_{A_2}$ & $i \cdot \boldsymbol{\chi}_{B_1} $ & $i \cdot \boldsymbol{\chi}_{B_2} $\\
	\end{tabular}\\
\end{table}

However, in case of finite magnetic-field calculations GIAOs are usually used. Here the proof, that real energy values are obtained if one component of the magnetic-field vector vanishes is similar, but the proof that the integrals are real is more involved. This can be seen in Appendix C.

\subsection{Complex entries in case of spin-orbit coupling}
Finally, we mention the case of relativistic quantum-chemical calculations with inclusion of spin-orbit coupling. This situation is very similar to that of a molecule in a magnetic field. In most methods with spin-orbit coupling \cite{CowanGriffin66, SaueViss03, Berning00} the expression of the Hamiltonian operator in second quantization equals the known representation from Eq.~(\ref{equation: Hamiltonian}). Then, analogously to the case of the presence of a finite magnetic field, the matrix elements $g$ and $h$ may be complex-valued and the CC equations have the same form as in Eq.~(\ref{equation: CC-equations}). The EOM-CCSD energy values are determined via the eigenvalues of $\bar{H}_\mathrm{P}$ as described in Section \ref{subsec: Coupled cluster}. The matrix $\bar{H}_\mathrm{P}$ has complex entries, which may lead to complex energy values. \\
Similar to finite magnetic-field calculations, a transformation to a real $H_\mathrm{FCI}$ matrix may exist in case of point-group or time-reversal symmetry. This has been, for example, shown for some quantum-chemical methods\cite{Viss96, Pitzer87} provided the molecular point group is $C_{2v}$, $D_2$ or one of its subgroups. Real energy values for the specified cases are in this way ensured, though in the absence of symmetry complex energy values are expected to be the normal case.

	\section{Concluding remarks} 
	Until now, complex energy values have rarely been observed  in CC calculations. This is not surprising, since we are showing in this paper that in standard CC calculations (i.e., those with real-valued Hamiltonian matrices) no complex energy values can occur for the ground state (see section \ref{subsec: Connection between}).\\
	In EOM-CC calculations, complex energy values are expected in the vicinity of conical intersections, as already mentioned\cite{Haettig2005} and observed \cite{Koehn2007, KMMK2017} in the literature. However, complex energy values only appear as long there is no constraint (e.g., due to symmetry) that ensures that the effective Hamiltonian matrix is non-defective. If complex energy values occur, they must be handled with care, since they come with  unwanted consequences. First of all they lead to a wrong dimension for the intersection seam (see Section \ref{subsec: Complex energy}). Furthermore, they lead to a wrong shape of the potential surface in the vicinity of the intersection (see Section \ref{subsec: Shape of}). Here we can mathematically deduce the root-like behaviour, which first was observed by Köhn and Tajti.\cite{Koehn2007} Along the way, we get evidence that the potential surface shows a linear dependence on the geometrical parameters near the intersection in the case of a non-defective matrix which is consistent with the results from Koch and co-workers.\cite{KMMK2017}
	One last unwanted consequence of complex energy values is the relatively large inaccuracy compared to the FCI solution. We have explained this finding theoretically and observed it in sample calculations for the 6-state-model.\\
	
	In the case of complex-valued entries in the matrix representation of the Hamiltonian, as they occur in finite magnetic-field CC and relativistic CC calculations with consideration of spin-orbit coupling, we have shown based on mathematical arguments that complex energy values can occur even if the state is well isolated and no conical intersection point lies nearby. By performing calculations for a $\mathrm{H_2O}$ molecule in a strong magnetic field we have confirmed this finding. \\
	Due to the established connection between FCI and CCSD energy values (see Section \ref{subsec: Connection between}) the appearing imaginary part in many cases can be considered as a kind of "numerical inaccuracy". Therefore the real part of the complex energy value provides a meaningful approximation to the exact energy value, as long as the CCSD method provides a good approximation to the FCI method and the imaginary part is sufficiently small.\\
	The fact that the previous literature did neither report complex energy values for finite magnetic-field CC calculations\cite{S15, HS17, HS19, HGS20} nor for CC calculations with inclusion of spin-orbit coupling\cite{WGW08, Liu18, Asthana19} is explained by our finding that symmetry might offer the possibility to transform the complex $\bar{H}_\mathrm{FCI}$ matrix into a real representation.

\section*{Acknowledgments}	

This paper is dedicated to Professor John Stanton on the occasion of his 60th birthday. One of the authors (J.G.) thanks him for more than 30 years of friendship and intense scientific collaborations which led to the development of the {\sc CFOUR} program package and about 90 joint publications.

The authors thank Professor Martin Hanke-Bourgeois (Johannes Gutenberg-Universit{\"a}t Mainz) for fruitful discussions concerning eigenvalue theory and acknowledge helpful discussions with Marios-Petros Kitsaras (Mainz), Dr. Simen Kvaal (University of Oslo), and Professor Lan Cheng (Johns Hopkins University). 

This work has been supported by the Deutsche Forschungsgemeinschaft via grant STO-1239/1-1.

	\begin{appendix}
		
\section{Mathematical proofs}
\label{Mathematical additions}
At this point we provide the mathematical proofs for some of the statements used in Section \ref{subsec: Mathematical Background}.

\begin{lemma}
	\label{reelle Koeff}
	Let $U(0)$ be a neighbourhood of $0$ in the complex plane.\\
	Let $f: U(0) \rightarrow \mathbb{C}$ be an analytical function in $U(0)$ for which holds:
	\begin{equation*}
	f(z)\in \mathbb{R}, \hspace*{0.5cm} \text{ for all } z \in U(0)\cap\mathbb{R}.
	\end{equation*}
	The coefficients of the power series of $f$ centered at $0$ are then real.
\end{lemma}
\begin{proof}
	Let $f(z)=p_0+p_{1}z+p_{2}z+p_{3}z^2+\cdots$ be the series expansion of the analytical function $f$ at the point $0$. For the coefficients $p_n$ it then holds that
	\begin{align*}
	p_n=\frac{f^{(n)}(0)}{n!},  \hspace*{0.5cm} \text{where } n\in \mathbb{N}.
	\end{align*}
	Let $z_0\in U(0)$ be real. Then, for $n=0$ the function $f$ fulfils
	\begin{align*}
	f^{(0)}(z_0):=f(z_0) \in \mathbb{R}.
	\end{align*}	
	according to the assumption above. Using mathematical induction starting from the fact that 
	\begin{align*}
	f^{(n)}(z_0)=\lim_{\substack{h\rightarrow 0 \\ h \in \mathbb{R}}} \left(\frac{f^{(n-1)}(z_0)-f^{(n-1)}(z_0+h)}{h}\right) \in \mathbb{R},
	\end{align*}
	yields the result that the values $f^{(n)}(0)$ are real for all $n\in \mathbb{N}$.
	Here the assumption $h\in \mathbb{R}$ is permitted, since the limit exists.
	Thus, all coefficients
	\begin{align*}
	p_n=\frac{f^{(n)}(0)}{n!}
	\end{align*}
	are real. 
\end{proof}

We continue by proving Theorem \ref{theorem: multiple eigenvalues}:
\begin{theoremWithoutNumber}
	Let $\lambda_i(\varepsilon)$ be an eigenvalue of the matrix $A(\varepsilon) \in \mathbb{C}(n \times n)$. Furthermore, let $0 < a,b \in \mathbb{R}$ exist, such that $\lambda_i(\varepsilon)$ takes a real value for all $\varepsilon \in [-a,0]$  and $\lambda_i(\varepsilon)$ takes a complex value for all  $\varepsilon \in (0,b]$. Then it holds:
	\begin{enumerate}[label=(\alph*)]
		\item In a neighbourhood of $\varepsilon=0$ the eigenvalue $\lambda_i(\varepsilon)$ can be represented by a branch of a Puiseux series.
		\item For $\varepsilon=0$ a multiple eigenvalue occurs.
	\end{enumerate} 
\end{theoremWithoutNumber}
\begin{proof}
	\begin{enumerate}[label=(\alph*)]
		\item 
		Assume that $\lambda_i(\varepsilon)$ has no representation as a branch of a Puiseux series as in Eq.~(\ref{equation: puiseux series of lambda}).
		Then, according to Theorem \ref{theorem: locale series}, $\lambda_i(\varepsilon)$ can be represented as a power series $\lambda_i(\varepsilon)=p_0+p_{1}\varepsilon+p_{2}\varepsilon^2+p_{3}\varepsilon^3+\cdots$, which converges for $\varepsilon \in (-r,r)$. \\
		Let $U_{r/2}(-\frac{r}{2})$ be the open circular disk with center $-\frac{r}{2}$ and radius $\frac{r}{2}$ in the complex plane. Let us define the analytical function
		\begin{align*}
		\Lambda(z): \hspace*{0.2cm} U_{r/2}\left(-\frac{r}{2}\right)\rightarrow \mathbb{C}, \hspace*{0.2cm}\\ \Lambda(z)=p_0+p_{1}z+p_{2}z^2+p_{3}z^3+\cdots,
		\end{align*} that coincides with $\lambda_i(\varepsilon)$ due to its definition on the interval $(-r,0)$.		
		According to the prerequisite it assumes only real values for all $z\in U_{r/2}(-\frac{r}{2})\cap \mathbb{R}$.
		Using Lemma \ref{reelle Koeff}, we conclude that the coefficients $p_{i}$ are real.\\
		For this reason, $\lambda_i(\varepsilon)$ for $\varepsilon \in (0,b]$ is real. This is a contradiction to the assumption. Thus statement $a)$ is proven.
		\item  
		According to Theorem \ref{theorem: locale series} the development of each simple eigenvalue $\lambda(\varepsilon)$ can be formulated by a power series. Since the eigenvalue $\lambda(0)$ is represented by a Puiseux series due to statement $a)$ of this theorem, a multiple eigenvalue has to occur for $\varepsilon=0$. 
	\end{enumerate}
\end{proof}
Let us finally prove Theorem \ref{theorem: real neigborhood}.
\begin{theoremWithoutNumber}
	Let $A(\varepsilon)$ be a matrix with only real entries for all $\varepsilon\in (-r,r)$, then:
	\begin{itemize}
		\item Let $\lambda_i(\varepsilon)$ be a complex eigenvalue of $A(\varepsilon)$, then the complex-conjugated value   $\lambda^*_i(\varepsilon)$ is also a complex eigenvalue of $A(\varepsilon)$.
			\item Let $\lambda_i(0)$ be a single real eigenvalue of $A(0)$.  Then a neighbourhood $U$ of $(0,\lambda_i(0))$ exists, in which $\lambda_i(\varepsilon)$ takes only real values.
	\end{itemize}
\end{theoremWithoutNumber}

\begin{proof}
	For a matrix with only real entries, the characteristic polynomial has only real coefficients. By applying the fundamental theorem of algebra we obtain the first result. \\
	For the second part let us assume that no neighbourhood of $(0,\lambda_i(0))$ exists, where $\lambda_i(\varepsilon)$ takes only real values. Then a sequence $\varepsilon_n$ with $\varepsilon_n \rightarrow 0$ exists, for which all eigenvalues $\lambda_i(\varepsilon_n)$ contain an imaginary part. \\
	According to the first statement the complex-conjugated values ${\lambda_i}^*(\varepsilon_n)$ are also eigenvalues of $A(\varepsilon)$. Since the eigenvalues depend on the parameter $\varepsilon$ in a continuous manner, the following holds: 
	\begin{align*}
	&{\lambda_i}(\varepsilon_n) \rightarrow {\lambda(0)}, \\
	&{\lambda_i}^*(\varepsilon_n) \rightarrow {\lambda(0)}, \hspace*{0.5cm}
	\text{where } {\lambda_i}^*(\varepsilon_n) \neq {\lambda_i}(\varepsilon_n). 
	\end{align*}
	Thus, in any neighborhood of $(0, \lambda (0))$ at least two eigenvalues for the same $ \varepsilon $ exist. This is a contradiction to Theorem \ref{theorem: locale series}, which proves this theorem.
\end{proof}

\section{Computational details for the 6-state model}
\label{sec: Computational details}
\small
Here, we provide computational details and additional results for the sample calculations on the 6-state model. In the matrix representation ${H}_\mathrm{FCI}$ of the Hamiltonian the counter-diagonal elements vanish due to the Slater-Condon rules. The first entry of ${H}_\mathrm{FCI}$ vanishes since the Hartree-Fock energy is subtracted from the diagonal elements. For the case of a real Hamiltonian the following matrix representation was chosen:
\setlength{\tabcolsep}{3pt} % Default value: 6pt
\renewcommand{\arraystretch}{0.8} % Default value: 1
%\small
\begin{align*}
H_\mathrm{FCI}=\left(
\scriptsize
\begin{matrix}
0  &  0.10  &  0.15  &  0.05  &  0.20     &    0
\\
0.10  &  0.50  &  -0.05  &  0.30  &   0  &  0.20
\\
0.15  &  -0.05  &  0.60   &     0  & -0.08  &  0.05
\\
0.05   & 0.30   &      0  &  0.70  &  -0.03  &  0.15
\\
0.20    &     0 &  -0.08  & -0.03  &  0.80  &  0.10
\\
0  &  0.20  &   0.05  &  0.15  &  0.10  &  1.50
\end{matrix}\right).
\end{align*}
A CC calculation provides here the amplitudes
\begin{align*}
&t_1= -0.2092, && t_2= -0.2579,\\
&t_3=0.0161, && t_4= -0.2486,
\end{align*}	
 which lead to the following EOM-CCSD matrix
 \setlength{\tabcolsep}{3pt} % Default value: 6pt
 \renewcommand{\arraystretch}{0.8} % Default value: 1
 \small
	\begin{align*}
	\bar{H}_\mathrm{P}=\left(
	\scriptsize\begin{matrix}
	-0.1085  &  0.1000  &  0.1500  &  0.0500  &  0.2000   \\
	0   & 0.4712  &  -0.0154  &  0.2589     &    0    \\
	0  &  -0.0366  &  0.6395  &       0  & -0.0389  \\ 
	0   & 0.2611   &      0   & 0.6605  & -0.0046   \\ 
	0    &    0  & -0.0411  & 0.0166  &  0.8288    \\
	\end{matrix}\right).
	\end{align*}
As example for a complex-valued Hamiltonian the following matrix representation was chosen:
\begin{widetext}
\setlength{\tabcolsep}{3pt} % Default value: 6pt
\renewcommand{\arraystretch}{0.8} % Default value: 1
\small
	\begin{align*}
	H_\mathrm{FCI}=
	\scriptsize
	\left(\begin{array}{cccccc}
	0 & 0.1 -0.1i  & 0.15+ 0.05i  &0.05-0.05i  & 0.2 -0.1i &0\\
	0.1+0.1i & 0.5 & 0.06 -0.03i & 0.03 -0.1i  & 0& 0.2 -0.1i\\
	0.15 -0.05i & 0.06+ 0.03i & 0.6  & 0& 0.03 -0.1i & 0.05 -0.05i\\
	0.05 + 0.05i & 0.03+ 0.1i & 0 & 0.7& 0.06 -0.03i & 0.15+ 0.05i\\
	0.2+ 0.1i & 0 & 0.03+ 0.1i & 0.06+ 0.03i & 0.8 & 0.1 -0.1i\\
	0& 0.2+ 0.1i & 0.05+ 0.05i & 0.15 -0.05i & 0.1+ 0.1i  & 1.5\\
	\end{array} \right)\\
	\end{align*}
%\small
\end{widetext}
which leads to the amplitudes
\begin{align*}
&t_1= -0.1539 - 0.1980i, && t_2= -0.1882 + 0.0643i,\\
&t_3=-0.0624 - 0.0489i, && t_4= -0.2109 - 0.0881i,
\end{align*} and to the EOM-CC matrix
\setlength{\tabcolsep}{3pt} % Default value: 6pt
\renewcommand{\arraystretch}{0.8} % Default value: 1
%\small
\begin{widetext}
\begin{align*}
H_\mathrm{P}=
\scriptsize
\left(\begin{array}{cccccc}
-0.1231 & 0.1000 -0.1000i  & 0.1500+ 0.0500i & 0.05-0.05i  & 0.2 -0.1i  \\
0& 0.4842+0.0079i & 0.0558+0.038i & 0.0164 - 0.0661i  & 0\\
0 & 0.0574+ 0.0109i & 0.6259+0.0004i & 0& 0.0436 -0.1339i\\
0& 0.0139+ 0.0749i & 0 & 0.6741-0.0004& 0.0642 -0.638i\\
0& 0 & 0.0461 + 0.1251i & -0.0008+ 0.0090i & 0.0015-0.0681i\\
\end{array} \right)\\
\end{align*}
\end{widetext}
The eigenvalues for the two examples are given in Table \ref{tab: table1} and Table \ref{tab: table2}.

\onecolumngrid
\vspace*{0.5cm}
\begin{table}[H]
	\caption{Comparison for the FCI and CCSD eigenvalues of the chosen example real Hamiltonian.}
	\begin{tabular}{c|c|c|c}
		\hline \\
		CCSD & FCI  & Difference between real part & Difference between imaginary part \\
		eigenvalues & eigenvalues &of CCSD eigenvalues & of CCSD eigenvalues\\
		& &and FCI eigenvalues&and FCI eigenvalues\\
		\hline
		$-0.1085$ & $-0.1085$ & $\num{8.24E-06}$ & $0$\\
		$0.2881$ & $0.2876$ &$\num{5.59E-04}$ &$0$ \\
		$0.6317$ & $0.6290$ & $\num{2.69E-03}$ & $0$\\
		$0.8401+0.0049i$ & $0.8269$ & $\num{1.32E-02}$ & $\num{4.93E-03}$\\
		$0.8401-0.0049i$ & $0.8601$ & $\num{-2.00E-02}$ & $\num{-4.93E-03}$\\
		-- & $1.6050$ &-- &-- 
			 \\ \hline
	\end{tabular}
	\label{tab: table1}
\end{table}
\begin{table}[H]
	\caption{Comparison for the FCI and CCSD eigenvalues of the chosen example complex Hamiltonian.}
	\begin{tabular}{c|c|c|c}
		\hline \\
		CCSD	& FCI  & Difference between real part & Difference between imaginary part \\
		eigenvalues & eigenvalues &of CCSD eigenvalues & of CCSD eigenvalues\\
		&&and FCI eigenvalue&and FCI eigenvalue\\
		\hline
		$-0.1232-\num{3.16E-05} i$ & $-0.1230$ & $\num{-1.67E-04}$ & $\num{-3.16E-05} $\\
		$0.4322+\num{4.61E-04}i$ & $0.4306$ & $\num{1.61E-03}$ & $\num{4.61E-04}$\\
		$0.5741+\num{6.90E-03}i$ & $0.5820$ & $\num{-7.97E-03}$ & $\num{6.90E-03}$ \\
		$0.6779+\num{1.98E-03}i$& $0.6796$ & $\num{-1.68E-03}$ & $\num{1.98E-03}$\\
		$0.9159-\num{9.34E-03}i$& $0.9133$ & $\num{2.51E-03}$ & $\num{-9.34E-03}$\\
		-- & $1.6175$ & --&--
					 \\ \hline

	\end{tabular}
	\label{tab: table2}	
\end{table}
\twocolumngrid

\section{Transformation to a real representation for H$_2$O in case of symmetry and GIAOs}
\label{appendix:GIAOTransformation}
In the following it is shown that the symmetry-inspired transformation to a real representation from Section \ref{subsec: Complex energy values for molecules in strong magnetic fields} is also valid with gauge-including atomic orbitals (GIAOs).\cite{London37,Tellgren12} GIAOs have the form
\begin{align}
\chi^{GIAO}=e^{- \frac{i}{2c} \vec{B} \times (\vec{R}_\nu-\vec{R}_0) \cdot \vec{r}} \chi(\vec{r})
\end{align} 
with $c$ as the velocity of light, $\vec{B}$ the magnetic-field vector, $\vec{R_{\nu}}$ the coordinates of $\nu$-th nucleus, $\vec{R}_0$ the gauge origin (in the following set to the origin of the coordinate system), $\chi$ the standard real basis function, and $\vec{r}$ the coordinates of the electron.\\

\textbf{GIAOs of the water molecule}\\
Let the water molecule be placed in the $yz$ plane as described in Section \ref{subsec: Complex energy values for molecules in strong magnetic fields} with
the oxygen atom at the origin of the coordinate system and $\pm y_H$ and $z_H$ the coordinates of the two hydrogen atoms. The (non-symmetry-adapted) GIAOs $\chi_{H_{1/2}}^{GIAO}$ of the hydrogen atoms are then given by:
\begin{align}
\chi_{H_{1/2}}^{GIAO}&= e^{- \frac{i}{2c} \left(\begin{psmallmatrix}
	B_x\\
	B_y\\
	B_z\\
	\end{psmallmatrix} \times \begin{psmallmatrix}
	0\\
	\pm y_H\\
	z_H\\
	\end{psmallmatrix}\right) \cdot \begin{psmallmatrix}
	x\\
	y\\
	z\\
	\end{psmallmatrix}} \chi_{H_{1/2}}.  \hspace*{0.5cm} 
\end{align}
 	A Taylor expansion of the exponential up to second order yields
\begin{align}
\chi_{H_{1/2}}^{GIAO}=&\chi_{H_{1/2}} - \frac{i}{2c} \pm y_H (B_x z - B_z x) \chi_{H_{1/2}}\nonumber \\
&  - \frac{i}{2c} z_H (B_y x -B_x y) \chi_{H_{1/2}}\nonumber\\
& -0.5 (\frac{1}{2c} y_H)^2 (\pm B_x z \mp B_z x)^2 \chi_{H_{1/2}}\nonumber \\
&- (\frac{1}{2c})^2 (\pm y_H) z_H (B_x z - B_z x) \nonumber\\ & \qquad\qquad (B_y x -B_x y) \chi_{H_{1/2}} \nonumber\\& + 0.5 (\frac{1}{2c} z_H)^2 (B_y x -B_x y)^2 \chi_{H_{1/2}}\nonumber\\ &+\cdots
\end{align}
Note that due to our choice of the coordinate system the GIAOs of the oxygen atom are identical to the corresponding AOs.

Symmetry adaptation then leads to
the following second-order expression for the symmetry-adapted hydrogen GIAOs:
\begin{align}
\chi_{H_{\pm}}^{GIAO}= &\chi_{H_{1}}^{GIAO} \pm \chi_{H_{2}}^{GIAO}\nonumber\\
= & (\chi_{H_{1}} \pm \chi_{H_{2}})\nonumber\\
 & - \frac{i}{2c} y_H (B_xz - B_zx) (\chi_{H_{1}}\mp \chi_{H_{2}})\nonumber\\
 & - \frac{i}{2c} z_H (B_y x -B_x y) (\chi_{H_{1}} \pm \chi_{H_{2}})\nonumber\\
  & -0.5 (\frac{1}{2c} y_H)^2 (- B_x z + B_zx)^2 (\chi_{H_{1}} \pm \chi_{H_{2}})\nonumber\\
  &- (\frac{1}{2c})^2 y_H z_H (B_x z - B_z x) \nonumber\\& \qquad\qquad (B_y x -B_x y) (\chi_{H_{1}} \mp \chi_{H_{2}}) \nonumber\\
 & - 0.5 (\frac{1}{2c} z_H)^2 (B_y x - B_x y)^2 (\chi_{H_{1}} \pm \chi_{H_{2}})
 \nonumber\\ &+\cdots
 \end{align}
\\

\textbf{Symmetry classification of real and imaginary part}\\
It is now rather straightforward to see that the real and imaginary terms in the expansion belong to different irreducible representations provided one magnetic-field component vanishes (see Table~\ref{table:giao}). For example, in case of $B_x=0$, the real contributions in the expansion are of $A_1$ and $B_2$ symmetry for AOs of $A_1$ and $B_2$ symmetry, while the imaginary terms are of $B_1$ and $A_2$ symmetry. For AOs of $B_1$ and $A_2$ symmetry, the situation is reversed and the real terms are of $B_1$ and $A_2$ symmetry, while the imaginary contributions are of $A_1 $ and $B_2$ symmetry. This observation suggests that the unitary transformation introduced in Section \ref{subsec: Complex energy values for molecules in strong magnetic fields} provides a consistent representation in which all real contributions are of $A_1$ and $B_2$ symmetry and all imaginary contributions are of $B_1$ and $A_2$ symmetry. The proof that this unitary transformation provides a real representation of all relevant one- and two-electron integrals as well of the Hamiltonian matrix is then analogous to the one given in Section \ref{subsec: Complex energy values for molecules in strong magnetic fields}. For the cases in which one of the other magnetic-field components vanishes, the proof can be carried out in an similar manner with grouping $A_1$ and $B_1$ as well as $B_2$ and $A_2$ symmetries together in case of $B_y=0$ and $A_1$ and $A_2$ as well $B_1$ and $B_2$ symmetries in the case of $B_z=0$.

The proof that in case of symmetry CC calculation for systems in finite magnetic fields can be carried out with real Hamiltonians always holds provided that for the given symmetry-adapted GIAOs the real and imaginary terms belong to different irreducible representations.
\begin{table}[H]
	\caption{Symmetry classification of the real and imaginary parts of the GIAOs for water depending on the irreducible representation of the underlying AO.}
	\label{table:giao}
	\begin{tabular}{lcccc} \\ \hline \vspace{-0.15cm}\\
&\multicolumn{4}{c}{symmetry of AO}\\ \cline{2-5}\vspace{-0.15cm}\\
			& A$_1$  & B$_1$ & B$_2$ & A$_2$ \\ \hline \vspace{-0.15cm}\\
	\multicolumn{4}{l}{a) $B_x$=0}\\
	real part  & A$_1$,B$_2$  & A$_2$, B$_1$ & A$_1$, B$_2$ & B$_1$,A$_2$ \\
		imag. part  & B$_1$, A$_2$  & A$_1$, B$_2$ & A$_2$, B$_1$ & A$_1$, B$_2$ \\
	\multicolumn{4}{l}{a) $B_y$=0}\\
	real part  & A$_1$,B$_1$  & A$_1$, B$_1$ & B$_2$, A$_2$ & B$_2$, A$_2$ \\
		imag. part & B$_2$, A$_2$  & B$_2$, A$_2$ & A$_1$, B$_1$ & A$_1$, B$_1$ \\
	\multicolumn{4}{l}{a) $B_z$=0}\\
	real part& A$_1$, A$_2$  & B$_1$, B$_2$ & B$_1$, B$_2$ & A$_1$, A$_2$ \\
		imag. part  & B$_1$, B$_2$  & A$_1$, A$_2$ & A$_1$, A$_2$ & B$_1$, B$_2$ \\\vspace{-0.15cm}\\ \hline
		\end{tabular}\\
\end{table}

	\end{appendix}
	\bibliography{appearance_of_complex_energy_values_in_cc_theory}

\begin{thebibliography}{55}%
\makeatletter
\providecommand \@ifxundefined [1]{%
 \@ifx{#1\undefined}
}%
\providecommand \@ifnum [1]{%
 \ifnum #1\expandafter \@firstoftwo
 \else \expandafter \@secondoftwo
 \fi
}%
\providecommand \@ifx [1]{%
 \ifx #1\expandafter \@firstoftwo
 \else \expandafter \@secondoftwo
 \fi
}%
\providecommand \natexlab [1]{#1}%
\providecommand \enquote  [1]{``#1''}%
\providecommand \bibnamefont  [1]{#1}%
\providecommand \bibfnamefont [1]{#1}%
\providecommand \citenamefont [1]{#1}%
\providecommand \href@noop [0]{\@secondoftwo}%
\providecommand \href [0]{\begingroup \@sanitize@url \@href}%
\providecommand \@href[1]{\@@startlink{#1}\@@href}%
\providecommand \@@href[1]{\endgroup#1\@@endlink}%
\providecommand \@sanitize@url [0]{\catcode `\\12\catcode `\$12\catcode
  `\&12\catcode `\#12\catcode `\^12\catcode `\_12\catcode `\%12\relax}%
\providecommand \@@startlink[1]{}%
\providecommand \@@endlink[0]{}%
\providecommand \url  [0]{\begingroup\@sanitize@url \@url }%
\providecommand \@url [1]{\endgroup\@href {#1}{\urlprefix }}%
\providecommand \urlprefix  [0]{URL }%
\providecommand \Eprint [0]{\href }%
\providecommand \doibase [0]{http://dx.doi.org/}%
\providecommand \selectlanguage [0]{\@gobble}%
\providecommand \bibinfo  [0]{\@secondoftwo}%
\providecommand \bibfield  [0]{\@secondoftwo}%
\providecommand \translation [1]{[#1]}%
\providecommand \BibitemOpen [0]{}%
\providecommand \bibitemStop [0]{}%
\providecommand \bibitemNoStop [0]{.\EOS\space}%
\providecommand \EOS [0]{\spacefactor3000\relax}%
\providecommand \BibitemShut  [1]{\csname bibitem#1\endcsname}%
\let\auto@bib@innerbib\@empty
%</preamble>
\bibitem [{\citenamefont {Shavitt}\ and\ \citenamefont
  {Bartlett}(2009)}]{SB09}%
  \BibitemOpen
  \bibfield  {author} {\bibinfo {author} {\bibfnamefont {I.}~\bibnamefont
  {Shavitt}}\ and\ \bibinfo {author} {\bibfnamefont {R.~J.}\ \bibnamefont
  {Bartlett}},\ }\href@noop {} {\emph {\bibinfo {title} {Many-Body Methods in
  Chemistry and Physics}}}\ (\bibinfo  {publisher} {Cambridge University
  Press},\ \bibinfo {address} {Cambridge},\ \bibinfo {year} {2009})\BibitemShut
  {NoStop}%
\bibitem [{\citenamefont {Emrich}(1981)}]{Emrich81}%
  \BibitemOpen
  \bibfield  {author} {\bibinfo {author} {\bibfnamefont {K.}~\bibnamefont
  {Emrich}},\ }\href@noop {} {\bibfield  {journal} {\bibinfo  {journal} {Nucl.
  Phys. A}\ }\textbf {\bibinfo {volume} {351}},\ \bibinfo {pages} {379}
  (\bibinfo {year} {1981})}\BibitemShut {NoStop}%
\bibitem [{\citenamefont {Stanton}\ and\ \citenamefont
  {Bartlett}(1993{\natexlab{a}})}]{Stanton93}%
  \BibitemOpen
  \bibfield  {author} {\bibinfo {author} {\bibfnamefont {J.~F.}\ \bibnamefont
  {Stanton}}\ and\ \bibinfo {author} {\bibfnamefont {R.~J.}\ \bibnamefont
  {Bartlett}},\ }\href@noop {} {\bibfield  {journal} {\bibinfo  {journal} {J.
  Chem. Phys.}\ }\textbf {\bibinfo {volume} {98}},\ \bibinfo {pages} {7029}
  (\bibinfo {year} {1993}{\natexlab{a}})}\BibitemShut {NoStop}%
\bibitem [{\citenamefont {Comeau}\ and\ \citenamefont
  {Bartlett}(1993)}]{Comeau93}%
  \BibitemOpen
  \bibfield  {author} {\bibinfo {author} {\bibfnamefont {D.~C.}\ \bibnamefont
  {Comeau}}\ and\ \bibinfo {author} {\bibfnamefont {R.~J.}\ \bibnamefont
  {Bartlett}},\ }\href@noop {} {\bibfield  {journal} {\bibinfo  {journal}
  {Chem. Phys. Lett.}\ }\textbf {\bibinfo {volume} {207}},\ \bibinfo {pages}
  {414} (\bibinfo {year} {1993})}\BibitemShut {NoStop}%
\bibitem [{\citenamefont {Rico}\ and\ \citenamefont
  {Head-Gordon}(1993)}]{Rico93}%
  \BibitemOpen
  \bibfield  {author} {\bibinfo {author} {\bibfnamefont {R.~J.}\ \bibnamefont
  {Rico}}\ and\ \bibinfo {author} {\bibfnamefont {M.}~\bibnamefont
  {Head-Gordon}},\ }\href@noop {} {\bibfield  {journal} {\bibinfo  {journal}
  {Chem. Phys. Lett.}\ }\textbf {\bibinfo {volume} {213}},\ \bibinfo {pages}
  {224} (\bibinfo {year} {1993})}\BibitemShut {NoStop}%
\bibitem [{\citenamefont {H{\"a}ttig}(2005)}]{Haettig2005}%
  \BibitemOpen
  \bibfield  {author} {\bibinfo {author} {\bibfnamefont {C.}~\bibnamefont
  {H{\"a}ttig}},\ }\href@noop {} {\bibfield  {journal} {\bibinfo  {journal}
  {Adv. Quant. Chem.}\ }\textbf {\bibinfo {volume} {50}},\ \bibinfo {pages}
  {37} (\bibinfo {year} {2005})}\BibitemShut {NoStop}%
\bibitem [{\citenamefont {K{\"o}hn}\ and\ \citenamefont
  {Tajti}(2007)}]{Koehn2007}%
  \BibitemOpen
  \bibfield  {author} {\bibinfo {author} {\bibfnamefont {A.}~\bibnamefont
  {K{\"o}hn}}\ and\ \bibinfo {author} {\bibfnamefont {A.}~\bibnamefont
  {Tajti}},\ }\href@noop {} {\bibfield  {journal} {\bibinfo  {journal} {J.
  Chem. Phys.}\ }\textbf {\bibinfo {volume} {127}},\ \bibinfo {pages} {044105}
  (\bibinfo {year} {2007})}\BibitemShut {NoStop}%
\bibitem [{\citenamefont {Kj{\o}nstad}\ \emph {et~al.}(2017)\citenamefont
  {Kj{\o}nstad}, \citenamefont {Myhre}, \citenamefont {Martinez},\ and\
  \citenamefont {Koch}}]{KMMK2017}%
  \BibitemOpen
  \bibfield  {author} {\bibinfo {author} {\bibfnamefont {E.~F.}\ \bibnamefont
  {Kj{\o}nstad}}, \bibinfo {author} {\bibfnamefont {R.~H.}\ \bibnamefont
  {Myhre}}, \bibinfo {author} {\bibfnamefont {T.~J.}\ \bibnamefont {Martinez}},
  \ and\ \bibinfo {author} {\bibfnamefont {H.}~\bibnamefont {Koch}},\
  }\href@noop {} {\bibfield  {journal} {\bibinfo  {journal} {J. Chem. Phys.}\
  }\textbf {\bibinfo {volume} {147}},\ \bibinfo {pages} {164105} (\bibinfo
  {year} {2017})}\BibitemShut {NoStop}%
\bibitem [{\citenamefont {Stopkowicz}\ \emph {et~al.}(2015)\citenamefont
  {Stopkowicz}, \citenamefont {Gauss}, \citenamefont {Lange}, \citenamefont
  {Tellgren},\ and\ \citenamefont {Helgaker}}]{S15}%
  \BibitemOpen
  \bibfield  {author} {\bibinfo {author} {\bibfnamefont {S.}~\bibnamefont
  {Stopkowicz}}, \bibinfo {author} {\bibfnamefont {J.}~\bibnamefont {Gauss}},
  \bibinfo {author} {\bibfnamefont {K.~K.}\ \bibnamefont {Lange}}, \bibinfo
  {author} {\bibfnamefont {E.~I.}\ \bibnamefont {Tellgren}}, \ and\ \bibinfo
  {author} {\bibfnamefont {T.}~\bibnamefont {Helgaker}},\ }\href@noop {}
  {\bibfield  {journal} {\bibinfo  {journal} {J. Chem. Phys.}\ }\textbf
  {\bibinfo {volume} {143}},\ \bibinfo {pages} {074110} (\bibinfo {year}
  {2015})}\BibitemShut {NoStop}%
\bibitem [{\citenamefont {Hampe}\ and\ \citenamefont
  {Stopkowicz}(2017)}]{HS17}%
  \BibitemOpen
  \bibfield  {author} {\bibinfo {author} {\bibfnamefont {F.}~\bibnamefont
  {Hampe}}\ and\ \bibinfo {author} {\bibfnamefont {S.}~\bibnamefont
  {Stopkowicz}},\ }\href@noop {} {\bibfield  {journal} {\bibinfo  {journal} {J.
  Chem. Phys.}\ }\textbf {\bibinfo {volume} {146}},\ \bibinfo {pages} {154105}
  (\bibinfo {year} {2017})}\BibitemShut {NoStop}%
\bibitem [{\citenamefont {Hampe}\ \emph {et~al.}(2020)\citenamefont {Hampe},
  \citenamefont {Gross},\ and\ \citenamefont {Stopkowicz}}]{HGS20}%
  \BibitemOpen
  \bibfield  {author} {\bibinfo {author} {\bibfnamefont {F.}~\bibnamefont
  {Hampe}}, \bibinfo {author} {\bibfnamefont {N.}~\bibnamefont {Gross}}, \ and\
  \bibinfo {author} {\bibfnamefont {S.}~\bibnamefont {Stopkowicz}},\
  }\href@noop {} {\bibfield  {journal} {\bibinfo  {journal} {Phys. Chem. Chem.
  Phys.}\ }\textbf {\bibinfo {volume} {22}},\ \bibinfo {pages} {23522}
  (\bibinfo {year} {2020})}\BibitemShut {NoStop}%
\bibitem [{\citenamefont {Visscher}\ \emph {et~al.}(1996)\citenamefont
  {Visscher}, \citenamefont {Lee},\ and\ \citenamefont {Dyall}}]{Visscher96}%
  \BibitemOpen
  \bibfield  {author} {\bibinfo {author} {\bibfnamefont {L.}~\bibnamefont
  {Visscher}}, \bibinfo {author} {\bibfnamefont {T.~L.}\ \bibnamefont {Lee}}, \
  and\ \bibinfo {author} {\bibfnamefont {K.~G.}\ \bibnamefont {Dyall}},\
  }\href@noop {} {\bibfield  {journal} {\bibinfo  {journal} {J. Chem. Phys.}\
  }\textbf {\bibinfo {volume} {105}},\ \bibinfo {pages} {8769} (\bibinfo {year}
  {1996})}\BibitemShut {NoStop}%
\bibitem [{\citenamefont {Wang}\ \emph {et~al.}(2008)\citenamefont {Wang},
  \citenamefont {Gauss},\ and\ \citenamefont {van Wüllen}}]{WGW08}%
  \BibitemOpen
  \bibfield  {author} {\bibinfo {author} {\bibfnamefont {F.}~\bibnamefont
  {Wang}}, \bibinfo {author} {\bibfnamefont {J.}~\bibnamefont {Gauss}}, \ and\
  \bibinfo {author} {\bibfnamefont {C.}~\bibnamefont {van Wüllen}},\
  }\href@noop {} {\bibfield  {journal} {\bibinfo  {journal} {J. Chem. Phys.}\
  }\textbf {\bibinfo {volume} {129}},\ \bibinfo {pages} {064113} (\bibinfo
  {year} {2008})}\BibitemShut {NoStop}%
\bibitem [{\citenamefont {Shee}\ \emph {et~al.}(2018)\citenamefont {Shee},
  \citenamefont {Saue}, \citenamefont {Visscher},\ and\ \citenamefont
  {Gomes}}]{Shee18}%
  \BibitemOpen
  \bibfield  {author} {\bibinfo {author} {\bibfnamefont {A.}~\bibnamefont
  {Shee}}, \bibinfo {author} {\bibfnamefont {T.}~\bibnamefont {Saue}}, \bibinfo
  {author} {\bibfnamefont {T.}~\bibnamefont {Visscher}}, \ and\ \bibinfo
  {author} {\bibfnamefont {A.~S.~P.}\ \bibnamefont {Gomes}},\ }\href@noop {}
  {\bibfield  {journal} {\bibinfo  {journal} {J. Chem. Phys.}\ }\textbf
  {\bibinfo {volume} {149}},\ \bibinfo {pages} {174113} (\bibinfo {year}
  {2018})}\BibitemShut {NoStop}%
\bibitem [{\citenamefont {Liu}\ \emph {et~al.}(2018{\natexlab{a}})\citenamefont
  {Liu}, \citenamefont {Shen}, \citenamefont {Asthana},\ and\ \citenamefont
  {Cheng}}]{Liu18}%
  \BibitemOpen
  \bibfield  {author} {\bibinfo {author} {\bibfnamefont {J.}~\bibnamefont
  {Liu}}, \bibinfo {author} {\bibfnamefont {Y.}~\bibnamefont {Shen}}, \bibinfo
  {author} {\bibfnamefont {A.}~\bibnamefont {Asthana}}, \ and\ \bibinfo
  {author} {\bibfnamefont {L.}~\bibnamefont {Cheng}},\ }\href@noop {}
  {\bibfield  {journal} {\bibinfo  {journal} {J. Chem. Phys.}\ }\textbf
  {\bibinfo {volume} {148}},\ \bibinfo {pages} {034106} (\bibinfo {year}
  {2018}{\natexlab{a}})}\BibitemShut {NoStop}%
\bibitem [{\citenamefont {Wilkinson}(1965)}]{Wilkinson1965}%
  \BibitemOpen
  \bibfield  {author} {\bibinfo {author} {\bibfnamefont {J.~H.}\ \bibnamefont
  {Wilkinson}},\ }\href@noop {} {\emph {\bibinfo {title} {The Algebraic
  Eigenvalue Problem}}},\ \bibinfo {edition} {revised.}\ ed.\ (\bibinfo
  {publisher} {Clarendon Press},\ \bibinfo {address} {Oxford},\ \bibinfo {year}
  {1965})\BibitemShut {NoStop}%
\bibitem [{\citenamefont {Kato}(1995)}]{Kato}%
  \BibitemOpen
  \bibfield  {author} {\bibinfo {author} {\bibfnamefont {T.}~\bibnamefont
  {Kato}},\ }\href@noop {} {\emph {\bibinfo {title} {Perturbation Theory for
  Linear Operators}}}\ (\bibinfo  {publisher} {Springer},\ \bibinfo {address}
  {Berlin Heidelberg},\ \bibinfo {year} {1995})\BibitemShut {NoStop}%
\bibitem [{\citenamefont {Wall}(2004)}]{Wall2004}%
  \BibitemOpen
  \bibfield  {author} {\bibinfo {author} {\bibfnamefont {C.~T.~C.}\
  \bibnamefont {Wall}},\ }\href@noop {} {\emph {\bibinfo {title} {Singular
  Points of Plane Curves}}}\ (\bibinfo  {publisher} {Cambridge University
  Press},\ \bibinfo {address} {Cambridge},\ \bibinfo {year} {2004})\BibitemShut
  {NoStop}%
\bibitem [{\citenamefont {Thomas}(2018)}]{Thomas18}%
  \BibitemOpen
  \bibfield  {author} {\bibinfo {author} {\bibfnamefont {S.}~\bibnamefont
  {Thomas}},\ }\emph {\bibinfo {title} {Komplexe Eigenwerte in der
  Equation-of-Motion Coupled-Cluster-Theorie}},\ \href@noop {} {Master's
  thesis},\ \bibinfo  {school} {Johannes Gutenberg-Universität Mainz}
  (\bibinfo {year} {2018})\BibitemShut {NoStop}%
\bibitem [{\citenamefont {Golub}\ and\ \citenamefont {Loan}(2013)}]{Golub2013}%
  \BibitemOpen
  \bibfield  {author} {\bibinfo {author} {\bibfnamefont {G.~H.}\ \bibnamefont
  {Golub}}\ and\ \bibinfo {author} {\bibfnamefont {C.~F.~V.}\ \bibnamefont
  {Loan}},\ }\href@noop {} {\emph {\bibinfo {title} {Matrix Computations}}},\
  \bibinfo {edition} {4th}\ ed.\ (\bibinfo  {publisher} {JHU Press},\ \bibinfo
  {address} {London},\ \bibinfo {year} {2013})\BibitemShut {NoStop}%
\bibitem [{\citenamefont {Dyall}\ and\ \citenamefont {{F{\ae}gri
  Jr.}}(2007)}]{Faegri}%
  \BibitemOpen
  \bibfield  {author} {\bibinfo {author} {\bibfnamefont {K.~G.}\ \bibnamefont
  {Dyall}}\ and\ \bibinfo {author} {\bibfnamefont {K.}~\bibnamefont {{F{\ae}gri
  Jr.}}},\ }\href@noop {} {\emph {\bibinfo {title} {Introduction to
  Relativistic Quantum Chemistry}}}\ (\bibinfo  {publisher} {Oxford University
  Press},\ \bibinfo {address} {New York},\ \bibinfo {year} {2007})\BibitemShut
  {NoStop}%
\bibitem [{\citenamefont {Hampe}\ and\ \citenamefont
  {Stopkowicz}(2019)}]{HS19}%
  \BibitemOpen
  \bibfield  {author} {\bibinfo {author} {\bibfnamefont {F.}~\bibnamefont
  {Hampe}}\ and\ \bibinfo {author} {\bibfnamefont {S.}~\bibnamefont
  {Stopkowicz}},\ }\href@noop {} {\bibfield  {journal} {\bibinfo  {journal} {J.
  Chem. Theory Comput.}\ }\textbf {\bibinfo {volume} {15}},\ \bibinfo {pages}
  {4036} (\bibinfo {year} {2019})}\BibitemShut {NoStop}%
\bibitem [{\citenamefont {Bartlett}\ and\ \citenamefont
  {Musia{\l}}(2007)}]{BM07}%
  \BibitemOpen
  \bibfield  {author} {\bibinfo {author} {\bibfnamefont {R.~J.}\ \bibnamefont
  {Bartlett}}\ and\ \bibinfo {author} {\bibfnamefont {M.}~\bibnamefont
  {Musia{\l}}},\ }\href@noop {} {\bibfield  {journal} {\bibinfo  {journal}
  {Rev. Mod. Phys.}\ }\textbf {\bibinfo {volume} {79}},\ \bibinfo {pages} {291}
  (\bibinfo {year} {2007})}\BibitemShut {NoStop}%
\bibitem [{\citenamefont {Schneider}(2009)}]{Schneider2009}%
  \BibitemOpen
  \bibfield  {author} {\bibinfo {author} {\bibfnamefont {R.}~\bibnamefont
  {Schneider}},\ }\href@noop {} {\bibfield  {journal} {\bibinfo  {journal}
  {Numerische Mathematik}\ }\textbf {\bibinfo {volume} {113}},\ \bibinfo
  {pages} {433} (\bibinfo {year} {2009})}\BibitemShut {NoStop}%
\bibitem [{\citenamefont {{Purvis III}}\ and\ \citenamefont
  {Bartlett}(1982)}]{Purvis82}%
  \BibitemOpen
  \bibfield  {author} {\bibinfo {author} {\bibfnamefont {G.~D.}\ \bibnamefont
  {{Purvis III}}}\ and\ \bibinfo {author} {\bibfnamefont {R.~J.}\ \bibnamefont
  {Bartlett}},\ }\href@noop {} {\bibfield  {journal} {\bibinfo  {journal} {J.
  Chem. Phys.}\ }\textbf {\bibinfo {volume} {76}},\ \bibinfo {pages} {1910}
  (\bibinfo {year} {1982})}\BibitemShut {NoStop}%
\bibitem [{\citenamefont {Rowe}(1968)}]{ROW68}%
  \BibitemOpen
  \bibfield  {author} {\bibinfo {author} {\bibfnamefont {D.~J.}\ \bibnamefont
  {Rowe}},\ }\href@noop {} {\bibfield  {journal} {\bibinfo  {journal} {Rev.
  Mod. Phys.}\ }\textbf {\bibinfo {volume} {40}},\ \bibinfo {pages} {153}
  (\bibinfo {year} {1968})}\BibitemShut {NoStop}%
\bibitem [{\citenamefont {Stanton}\ and\ \citenamefont
  {Bartlett}(1993{\natexlab{b}})}]{Stanton1993}%
  \BibitemOpen
  \bibfield  {author} {\bibinfo {author} {\bibfnamefont {J.~F.}\ \bibnamefont
  {Stanton}}\ and\ \bibinfo {author} {\bibfnamefont {R.~J.}\ \bibnamefont
  {Bartlett}},\ }\href@noop {} {\bibfield  {journal} {\bibinfo  {journal} {J.
  Chem. Phys.}\ }\textbf {\bibinfo {volume} {98}},\ \bibinfo {pages} {7029}
  (\bibinfo {year} {1993}{\natexlab{b}})}\BibitemShut {NoStop}%
\bibitem [{\citenamefont {Kowalski}\ and\ \citenamefont
  {Piecuch}(2001)}]{eomccsdt1}%
  \BibitemOpen
  \bibfield  {author} {\bibinfo {author} {\bibfnamefont {K.}~\bibnamefont
  {Kowalski}}\ and\ \bibinfo {author} {\bibfnamefont {P.}~\bibnamefont
  {Piecuch}},\ }\href@noop {} {\bibfield  {journal} {\bibinfo  {journal} {J.
  Chem. Phys.}\ }\textbf {\bibinfo {volume} {115}},\ \bibinfo {pages} {643}
  (\bibinfo {year} {2001})}\BibitemShut {NoStop}%
\bibitem [{\citenamefont {Kucharski}\ \emph {et~al.}(2001)\citenamefont
  {Kucharski}, \citenamefont {W{\l}och}, \citenamefont {Musia{\l}},\ and\
  \citenamefont {Bartlett}}]{eomccsdt2}%
  \BibitemOpen
  \bibfield  {author} {\bibinfo {author} {\bibfnamefont {S.~A.}\ \bibnamefont
  {Kucharski}}, \bibinfo {author} {\bibfnamefont {M.}~\bibnamefont {W{\l}och}},
  \bibinfo {author} {\bibfnamefont {M.}~\bibnamefont {Musia{\l}}}, \ and\
  \bibinfo {author} {\bibfnamefont {R.~J.}\ \bibnamefont {Bartlett}},\
  }\href@noop {} {\bibfield  {journal} {\bibinfo  {journal} {J. Chem. Phys.}\
  }\textbf {\bibinfo {volume} {115}},\ \bibinfo {pages} {8263} (\bibinfo {year}
  {2001})}\BibitemShut {NoStop}%
\bibitem [{\citenamefont {Bomble}\ \emph {et~al.}(2004)\citenamefont {Bomble},
  \citenamefont {Sattelmeyer}, \citenamefont {Stanton},\ and\ \citenamefont
  {Gauss}}]{eomccsdt3}%
  \BibitemOpen
  \bibfield  {author} {\bibinfo {author} {\bibfnamefont {Y.~J.}\ \bibnamefont
  {Bomble}}, \bibinfo {author} {\bibfnamefont {K.~W.}\ \bibnamefont
  {Sattelmeyer}}, \bibinfo {author} {\bibfnamefont {J.~F.}\ \bibnamefont
  {Stanton}}, \ and\ \bibinfo {author} {\bibfnamefont {J.}~\bibnamefont
  {Gauss}},\ }\href@noop {} {\bibfield  {journal} {\bibinfo  {journal} {J.
  Chem. Phys.}\ }\textbf {\bibinfo {volume} {121}},\ \bibinfo {pages} {5236}
  (\bibinfo {year} {2004})}\BibitemShut {NoStop}%
\bibitem [{\citenamefont {K\'{a}llay}\ and\ \citenamefont
  {Gauss}(2004)}]{eomccsdtq}%
  \BibitemOpen
  \bibfield  {author} {\bibinfo {author} {\bibfnamefont {M.}~\bibnamefont
  {K\'{a}llay}}\ and\ \bibinfo {author} {\bibfnamefont {J.}~\bibnamefont
  {Gauss}},\ }\href@noop {} {\bibfield  {journal} {\bibinfo  {journal} {J.
  Chem. Phys.}\ }\textbf {\bibinfo {volume} {121}},\ \bibinfo {pages} {9257}
  (\bibinfo {year} {2004})}\BibitemShut {NoStop}%
\bibitem [{\citenamefont {Yarkony}(1996)}]{Yarkony96}%
  \BibitemOpen
  \bibfield  {author} {\bibinfo {author} {\bibfnamefont {D.~R.}\ \bibnamefont
  {Yarkony}},\ }\href@noop {} {\bibfield  {journal} {\bibinfo  {journal} {Rev.
  Mod. Phys.}\ }\textbf {\bibinfo {volume} {68}},\ \bibinfo {pages} {985}
  (\bibinfo {year} {1996})}\BibitemShut {NoStop}%
\bibitem [{\citenamefont {Matsika}\ and\ \citenamefont
  {Yarkony}(2001)}]{Yarkony2001}%
  \BibitemOpen
  \bibfield  {author} {\bibinfo {author} {\bibfnamefont {S.}~\bibnamefont
  {Matsika}}\ and\ \bibinfo {author} {\bibfnamefont {D.~R.}\ \bibnamefont
  {Yarkony}},\ }\href@noop {} {\bibfield  {journal} {\bibinfo  {journal} {J.
  Chem. Phys.}\ }\textbf {\bibinfo {volume} {115}},\ \bibinfo {pages} {2038}
  (\bibinfo {year} {2001})}\BibitemShut {NoStop}%
\bibitem [{\citenamefont {Truhlar}\ and\ \citenamefont
  {Mead}(2003)}]{Truhlar03}%
  \BibitemOpen
  \bibfield  {author} {\bibinfo {author} {\bibfnamefont {D.~G.}\ \bibnamefont
  {Truhlar}}\ and\ \bibinfo {author} {\bibfnamefont {C.~A.}\ \bibnamefont
  {Mead}},\ }\href@noop {} {\bibfield  {journal} {\bibinfo  {journal} {Phys.
  Rev. A}\ }\textbf {\bibinfo {volume} {68}},\ \bibinfo {pages} {032501}
  (\bibinfo {year} {2003})}\BibitemShut {NoStop}%
\bibitem [{\citenamefont {Zhu}\ and\ \citenamefont
  {Yarkony}(2016)}]{Yarkony16}%
  \BibitemOpen
  \bibfield  {author} {\bibinfo {author} {\bibfnamefont {X.}~\bibnamefont
  {Zhu}}\ and\ \bibinfo {author} {\bibfnamefont {D.~R.}\ \bibnamefont
  {Yarkony}},\ }\href {\doibase 10.1080/00268976.2016.1170218} {\bibfield
  {journal} {\bibinfo  {journal} {Mol. Phys.}\ }\textbf {\bibinfo {volume}
  {114}},\ \bibinfo {pages} {1983} (\bibinfo {year} {2016})}\BibitemShut
  {NoStop}%
\bibitem [{\citenamefont {von Neuman}\ and\ \citenamefont
  {Wigner}(1929)}]{Neumann29}%
  \BibitemOpen
  \bibfield  {author} {\bibinfo {author} {\bibfnamefont {J.}~\bibnamefont {von
  Neuman}}\ and\ \bibinfo {author} {\bibfnamefont {E.}~\bibnamefont {Wigner}},\
  }\href@noop {} {\bibfield  {journal} {\bibinfo  {journal} {Physik. Z.}\
  }\textbf {\bibinfo {volume} {30}},\ \bibinfo {pages} {467} (\bibinfo {year}
  {1929})}\BibitemShut {NoStop}%
\bibitem [{\citenamefont {Keller}(2008)}]{Keller2008}%
  \BibitemOpen
  \bibfield  {author} {\bibinfo {author} {\bibfnamefont {J.}~\bibnamefont
  {Keller}},\ }\href@noop {} {\bibfield  {journal} {\bibinfo  {journal} {Linear
  Algebra \& its Applications}\ }\textbf {\bibinfo {volume} {429}},\ \bibinfo
  {pages} {2209} (\bibinfo {year} {2008})}\BibitemShut {NoStop}%
\bibitem [{\citenamefont {Teller}(1937)}]{Teller37}%
  \BibitemOpen
  \bibfield  {author} {\bibinfo {author} {\bibfnamefont {E.}~\bibnamefont
  {Teller}},\ }\href@noop {} {\bibfield  {journal} {\bibinfo  {journal} {J.
  Chem. Phys.}\ }\textbf {\bibinfo {volume} {42}},\ \bibinfo {pages} {109}
  (\bibinfo {year} {1937})}\BibitemShut {NoStop}%
\bibitem [{\citenamefont {Kj{\o}nstad}\ and\ \citenamefont
  {Koch}(2017)}]{KK2017}%
  \BibitemOpen
  \bibfield  {author} {\bibinfo {author} {\bibfnamefont {E.~F.}\ \bibnamefont
  {Kj{\o}nstad}}\ and\ \bibinfo {author} {\bibfnamefont {H.}~\bibnamefont
  {Koch}},\ }\href@noop {} {\bibfield  {journal} {\bibinfo  {journal} {J. Phys.
  Chem. Lett.}\ }\textbf {\bibinfo {volume} {8}},\ \bibinfo {pages} {4801}
  (\bibinfo {year} {2017})}\BibitemShut {NoStop}%
\bibitem [{\citenamefont {Schirmer}(1982)}]{adc2}%
  \BibitemOpen
  \bibfield  {author} {\bibinfo {author} {\bibfnamefont {J.}~\bibnamefont
  {Schirmer}},\ }\href@noop {} {\bibfield  {journal} {\bibinfo  {journal}
  {Phys. Rev. A}\ }\textbf {\bibinfo {volume} {26}},\ \bibinfo {pages} {2395}
  (\bibinfo {year} {1982})}\BibitemShut {NoStop}%
\bibitem [{\citenamefont {Watts}\ \emph {et~al.}(1989)\citenamefont {Watts},
  \citenamefont {Trucks},\ and\ \citenamefont {Bartlett}}]{ucc4}%
  \BibitemOpen
  \bibfield  {author} {\bibinfo {author} {\bibfnamefont {J.~D.}\ \bibnamefont
  {Watts}}, \bibinfo {author} {\bibfnamefont {G.~W.}\ \bibnamefont {Trucks}}, \
  and\ \bibinfo {author} {\bibfnamefont {R.~J.}\ \bibnamefont {Bartlett}},\
  }\href@noop {} {\bibfield  {journal} {\bibinfo  {journal} {Chem. Phys.
  Lett.}\ }\textbf {\bibinfo {volume} {157}},\ \bibinfo {pages} {359} (\bibinfo
  {year} {1989})}\BibitemShut {NoStop}%
\bibitem [{\citenamefont {Liu}\ \emph {et~al.}(2018{\natexlab{b}})\citenamefont
  {Liu}, \citenamefont {Asthana}, \citenamefont {Cheng},\ and\ \citenamefont
  {Mukherjee}}]{ucc}%
  \BibitemOpen
  \bibfield  {author} {\bibinfo {author} {\bibfnamefont {J.}~\bibnamefont
  {Liu}}, \bibinfo {author} {\bibfnamefont {A.}~\bibnamefont {Asthana}},
  \bibinfo {author} {\bibfnamefont {L.}~\bibnamefont {Cheng}}, \ and\ \bibinfo
  {author} {\bibfnamefont {D.}~\bibnamefont {Mukherjee}},\ }\href@noop {}
  {\bibfield  {journal} {\bibinfo  {journal} {J. Chem. Phys.}\ }\textbf
  {\bibinfo {volume} {148}},\ \bibinfo {pages} {244110} (\bibinfo {year}
  {2018}{\natexlab{b}})}\BibitemShut {NoStop}%
\bibitem [{\citenamefont {Grazioli}\ and\ \citenamefont
  {Stopkowicz}(2021)}]{ucclaura}%
  \BibitemOpen
  \bibfield  {author} {\bibinfo {author} {\bibfnamefont {L.}~\bibnamefont
  {Grazioli}}\ and\ \bibinfo {author} {\bibfnamefont {S.}~\bibnamefont
  {Stopkowicz}},\ }\href@noop {} {\enquote {\bibinfo {title} {Unitary coupled
  cluster theory for atoms and molecules in strong magnetic fields},}\ }
  (\bibinfo {year} {2021}),\ \bibinfo {note} {{i}n preparation}\BibitemShut
  {NoStop}%
\bibitem [{\citenamefont {Cowan}\ and\ \citenamefont
  {Griffin}(1976)}]{CowanGriffin66}%
  \BibitemOpen
  \bibfield  {author} {\bibinfo {author} {\bibfnamefont {R.~D.}\ \bibnamefont
  {Cowan}}\ and\ \bibinfo {author} {\bibfnamefont {D.}~\bibnamefont
  {Griffin}},\ }\href@noop {} {\bibfield  {journal} {\bibinfo  {journal} {J.
  Opt. Soc. Am.}\ }\textbf {\bibinfo {volume} {66}},\ \bibinfo {pages} {1010}
  (\bibinfo {year} {1976})}\BibitemShut {NoStop}%
\bibitem [{\citenamefont {Saue}\ and\ \citenamefont
  {Visscher}(2003)}]{SaueViss03}%
  \BibitemOpen
  \bibfield  {author} {\bibinfo {author} {\bibfnamefont {T.}~\bibnamefont
  {Saue}}\ and\ \bibinfo {author} {\bibfnamefont {L.}~\bibnamefont
  {Visscher}},\ }in\ \href@noop {} {\emph {\bibinfo {booktitle} {Theoretical
  Chemistry and Physics of Heavy and Superheavy Elements}}},\ \bibinfo {editor}
  {edited by\ \bibinfo {editor} {\bibfnamefont {U.}~\bibnamefont {Kaldor}}\
  and\ \bibinfo {editor} {\bibfnamefont {S.}~\bibnamefont {Wilson}}}\ (\bibinfo
   {publisher} {Kluwer Academic Publishers},\ \bibinfo {address} {Dordrecht},\
  \bibinfo {year} {2003})\ p.\ \bibinfo {pages} {211}\BibitemShut {NoStop}%
\bibitem [{\citenamefont {Berning}\ \emph {et~al.}(2000)\citenamefont
  {Berning}, \citenamefont {Schweizer}, \citenamefont {Werner}, \citenamefont
  {Knowles},\ and\ \citenamefont {Palmieri}}]{Berning00}%
  \BibitemOpen
  \bibfield  {author} {\bibinfo {author} {\bibfnamefont {A.}~\bibnamefont
  {Berning}}, \bibinfo {author} {\bibfnamefont {M.}~\bibnamefont {Schweizer}},
  \bibinfo {author} {\bibfnamefont {H.-J.}\ \bibnamefont {Werner}}, \bibinfo
  {author} {\bibfnamefont {P.~J.}\ \bibnamefont {Knowles}}, \ and\ \bibinfo
  {author} {\bibfnamefont {P.}~\bibnamefont {Palmieri}},\ }\href@noop {}
  {\bibfield  {journal} {\bibinfo  {journal} {Mol. Phys.}\ }\textbf {\bibinfo
  {volume} {98}},\ \bibinfo {pages} {1823} (\bibinfo {year}
  {2000})}\BibitemShut {NoStop}%
\bibitem [{\citenamefont {Kendall}\ and\ \citenamefont
  {Dunning~Jr.}(1992)}]{Kendall92}%
  \BibitemOpen
  \bibfield  {author} {\bibinfo {author} {\bibfnamefont {R.~A.}\ \bibnamefont
  {Kendall}}\ and\ \bibinfo {author} {\bibfnamefont {T.~H.}\ \bibnamefont
  {Dunning~Jr.}},\ }\href@noop {} {\bibfield  {journal} {\bibinfo  {journal}
  {J. Chem. Phys.}\ }\textbf {\bibinfo {volume} {96}},\ \bibinfo {pages} {6796}
  (\bibinfo {year} {1992})}\BibitemShut {NoStop}%
\bibitem [{\citenamefont {London}(1937)}]{London37}%
  \BibitemOpen
  \bibfield  {author} {\bibinfo {author} {\bibfnamefont {F.}~\bibnamefont
  {London}},\ }\href@noop {} {\bibfield  {journal} {\bibinfo  {journal} {J.
  Phys. Radium}\ }\textbf {\bibinfo {volume} {8}},\ \bibinfo {pages} {397}
  (\bibinfo {year} {1937})}\BibitemShut {NoStop}%
\bibitem [{\citenamefont {Tellgren}\ \emph {et~al.}(2012)\citenamefont
  {Tellgren}, \citenamefont {Reine},\ and\ \citenamefont
  {Helgaker}}]{Tellgren12}%
  \BibitemOpen
  \bibfield  {author} {\bibinfo {author} {\bibfnamefont {E.~I.}\ \bibnamefont
  {Tellgren}}, \bibinfo {author} {\bibfnamefont {S.~S.}\ \bibnamefont {Reine}},
  \ and\ \bibinfo {author} {\bibfnamefont {T.}~\bibnamefont {Helgaker}},\
  }\href@noop {} {\bibfield  {journal} {\bibinfo  {journal} {Phys. Chem. Chem.
  Phys.}\ }\textbf {\bibinfo {volume} {14}},\ \bibinfo {pages} {9492} (\bibinfo
  {year} {2012})}\BibitemShut {NoStop}%
\bibitem [{\citenamefont {Hampe}\ \emph {et~al.}()\citenamefont {Hampe},
  \citenamefont {Stopkowicz}, \citenamefont {Gross},\ and\ \citenamefont
  {Kitsaras}}]{qcumbre}%
  \BibitemOpen
  \bibfield  {author} {\bibinfo {author} {\bibfnamefont {F.}~\bibnamefont
  {Hampe}}, \bibinfo {author} {\bibfnamefont {S.}~\bibnamefont {Stopkowicz}},
  \bibinfo {author} {\bibfnamefont {N.}~\bibnamefont {Gross}}, \ and\ \bibinfo
  {author} {\bibfnamefont {M.-P.}\ \bibnamefont {Kitsaras}},\ }\href@noop {}
  {\enquote {\bibinfo {title} {{QCUMBRE}, {Q}uantum8-{C}hemical {U}tility
  enabling {M}agnetic-field dependent investigations {B}enefiting from
  {R}igorous {E}lectron-correlation treatment},}\ }\bibinfo {note}
  {{q}cumbre.org}\BibitemShut {NoStop}%
\bibitem [{\citenamefont {Tellgren}\ \emph {et~al.}(2008)\citenamefont
  {Tellgren}, \citenamefont {Soncini},\ and\ \citenamefont
  {Helgaker}}]{londonpap}%
  \BibitemOpen
  \bibfield  {author} {\bibinfo {author} {\bibfnamefont {E.~I.}\ \bibnamefont
  {Tellgren}}, \bibinfo {author} {\bibfnamefont {A.}~\bibnamefont {Soncini}}, \
  and\ \bibinfo {author} {\bibfnamefont {T.}~\bibnamefont {Helgaker}},\ }\href
  {http://aip.scitation.org/doi/10.1063/1.2996525} {\bibfield  {journal}
  {\bibinfo  {journal} {J. Chem. Phys.}\ }\textbf {\bibinfo {volume} {129}},\
  \bibinfo {pages} {154114} (\bibinfo {year} {2008})}\BibitemShut {NoStop}%
\bibitem [{\citenamefont {Tellgren}\ \emph {et~al.}()\citenamefont {Tellgren},
  \citenamefont {Helgaker}, \citenamefont {Soncini}, \citenamefont {Lange},
  \citenamefont {Teale}, \citenamefont {Ekström}, \citenamefont {Stopkowicz},
  \citenamefont {Austad},\ and\ \citenamefont {Sen}}]{london}%
  \BibitemOpen
  \bibfield  {author} {\bibinfo {author} {\bibfnamefont {E.~I.}\ \bibnamefont
  {Tellgren}}, \bibinfo {author} {\bibfnamefont {T.}~\bibnamefont {Helgaker}},
  \bibinfo {author} {\bibfnamefont {A.}~\bibnamefont {Soncini}}, \bibinfo
  {author} {\bibfnamefont {K.~K.}\ \bibnamefont {Lange}}, \bibinfo {author}
  {\bibfnamefont {A.~M.}\ \bibnamefont {Teale}}, \bibinfo {author}
  {\bibfnamefont {U.}~\bibnamefont {Ekström}}, \bibinfo {author}
  {\bibfnamefont {S.}~\bibnamefont {Stopkowicz}}, \bibinfo {author}
  {\bibfnamefont {J.~H.}\ \bibnamefont {Austad}}, \ and\ \bibinfo {author}
  {\bibfnamefont {S.}~\bibnamefont {Sen}},\ }\href@noop {} {\enquote {\bibinfo
  {title} {{LONDON}, a quantum-chemistry program for plane-wave/{GTO} hybrid
  basis sets and finite magnetic field calculations},}\ }\bibinfo {note}
  {{l}ondonprogram.org}\BibitemShut {NoStop}%
\bibitem [{\citenamefont {Pitzer}\ and\ \citenamefont
  {Winter}(1987)}]{Pitzer87}%
  \BibitemOpen
  \bibfield  {author} {\bibinfo {author} {\bibfnamefont {R.~M.}\ \bibnamefont
  {Pitzer}}\ and\ \bibinfo {author} {\bibfnamefont {N.~W.}\ \bibnamefont
  {Winter}},\ }\href@noop {} {\bibfield  {journal} {\bibinfo  {journal} {J.
  Phys. Chem.}\ }\textbf {\bibinfo {volume} {92}},\ \bibinfo {pages} {3061}
  (\bibinfo {year} {1987})}\BibitemShut {NoStop}%
\bibitem [{\citenamefont {Visscher}(1996)}]{Viss96}%
  \BibitemOpen
  \bibfield  {author} {\bibinfo {author} {\bibfnamefont {L.}~\bibnamefont
  {Visscher}},\ }\href@noop {} {\bibfield  {journal} {\bibinfo  {journal}
  {Chem. Phys. Lett.}\ }\textbf {\bibinfo {volume} {253}},\ \bibinfo {pages}
  {20} (\bibinfo {year} {1996})}\BibitemShut {NoStop}%
\bibitem [{\citenamefont {Asthana}\ \emph {et~al.}(2019)\citenamefont
  {Asthana}, \citenamefont {Liu},\ and\ \citenamefont {Cheng}}]{Asthana19}%
  \BibitemOpen
  \bibfield  {author} {\bibinfo {author} {\bibfnamefont {A.}~\bibnamefont
  {Asthana}}, \bibinfo {author} {\bibfnamefont {J.}~\bibnamefont {Liu}}, \ and\
  \bibinfo {author} {\bibfnamefont {L.}~\bibnamefont {Cheng}},\ }\href@noop {}
  {\bibfield  {journal} {\bibinfo  {journal} {J. Chem. Phys.}\ }\textbf
  {\bibinfo {volume} {150}},\ \bibinfo {pages} {0074102} (\bibinfo {year}
  {2019})}\BibitemShut {NoStop}%
\end{thebibliography}%
\end{document}